\documentclass[a4paper,11pt]{article}
\pdfoutput=1 

\pdfoutput=1
\usepackage{slashed}
\usepackage{makecell}
\usepackage{multirow}
\usepackage{graphicx,array}
\usepackage{booktabs}
\usepackage{setspace}
\usepackage{amstext}
\usepackage{amssymb}
\usepackage{mathtools}
\usepackage[T1]{fontenc} 
\usepackage{xcolor}
\usepackage{changepage}
\usepackage{arydshln}
\usepackage[makeroom]{cancel}
\usepackage{booktabs}
\usepackage{jheppub} 

\definecolor{orange}{rgb}{1,0.5,0}
\definecolor{brown}{rgb}{0.59, 0.29, 0.0}

\definecolor{note_fontcolor}{rgb}{0.80078125, 0.80078125, 0.80078125}

\def\beq{\begin{equation}}
\def\eeq{\end{equation}}
\def\bea{\begin{eqnarray}}
\def\eea{\end{eqnarray}}

\makeatletter
\newcommand*{\Relbarfill@}{\arrowfill@\Relbar\Relbar\Relbar}
\newcommand*{\xeq}[2][]{\ext@arrow 0055\Relbarfill@{#1}{#2}}
\makeatother

\newcommand{\bra}[1]{\left\langle #1 \right|}
\newcommand{\ket}[1]{\left| #1 \right\rangle}

\newcommand{\colvec}[1]{\left(\begin{array}{c}#1\end{array}\right)}

\newcommand{\colmatt}[1]{\left(\begin{array}{cc}#1\end{array}\right)}

\newcommand{\bk}[1]{\left<#1\right>}
\newcommand{\bks}[1]{\left[#1\right]}
\newcommand{\bmk}[3]{\left<#1|#2|#3\right]}

\newcommand{\sket}[1]{\left|#1\right]}
\newcommand{\sbra}[1]{\left[#1\right|}

\newcommand{\mypm}{\,\scriptstyle{\pm}\textstyle \,}

\def\s{\sigma}

\def\lpar#1#2#3#4{\rlap{\raise#3\hbox{$\hskip#4#1\left\{\mbox{\phantom{\rule[0mm]{0mm}{#2}}}\right.$}}}
\def\rpar#1#2#3#4{\rlap{\raise#3\hbox{$\hskip#4\left\}#1\mbox{\phantom{\rule[0mm]{0mm}{#2}}}\right.$}}}
\renewcommand{\subsubsection}[1]{\addtocounter{subsubsection}{1}
\par\nobreak
\medskip
\nobreak
\noindent{\it \thesubsubsection.  #1 }
\par\nobreak\medskip\nobreak}

\title{\boldmath  
Scattering Amplitudes for Monopoles: \\
Pairwise Little Group and Pairwise Helicity}

\author[a]{Csaba Cs\'aki,}
\author[a]{Sungwoo Hong,}
\author[b]{Yuri Shirman,}
\author[c,d]{Ofri Telem}
\author[e]{John Terning,}
\author[b]{and Michael Waterbury}

\affiliation[a]{Department of Physics, 
LEPP, Cornell University, Ithaca, NY 14853, USA}
\affiliation[b]{Department of Physics $\&$ Astronomy, University of California, Irvine, CA 92697, USA}
\affiliation[c]{Theory Group, Lawrence Berkeley National Laboratory, Berkeley, CA 94720, USA}
\affiliation[d]{Berkeley Center for Theoretical Physics, University of California, Berkeley, CA 94720, USA}
\affiliation[e]{
Department of Physics, University of California, Davis, CA 95616, USA}

\abstract{On-shell methods are particularly suited for exploring the scattering of electrically and magnetically charged objects, for which there is no local and Lorentz invariant Lagrangian description. In this paper we show how to construct a Lorentz-invariant $S$-matrix for the scattering of electrically and magnetically charged particles, without ever having to refer to a Dirac string.
A key ingredient is a revision of our fundamental understanding of multi-particle representations of the Poincar\'e group. Surprisingly, the asymptotic states for electric-magnetic scattering transform with an additional little group phase, associated with \textit{pairs} of electrically and magnetically charged particles. The corresponding ``pairwise helicity'' is identified with the quantized ``cross product'' of charges, $e_1 g_2 - e_2 g_1$, for every charge-monopole pair, and represents the extra angular momentum stored in the asymptotic electromagnetic field. We define a new kind of pairwise spinor-helicity variable, which serves as an additional building block for electric-magnetic scattering amplitudes. We then construct the most general 3-point $S$-matrix elements, as well as the full partial wave decomposition for the $2\to 2$ fermion-monopole $S$-matrix. In particular, we derive the famous helicity flip in the lowest partial wave as a simple consequence of a generalized spin-helicity selection rule, as well as the full angular dependence for the higher partial waves. Our construction provides a significant new achievement for the on-shell program, succeeding where the Lagrangian description has so far failed.}

\begin{document}
\maketitle
\flushbottom

\section{Introduction}

Unitary representations of the Poincar\'e group, classified by Wigner~\cite{Wigner:1939cj} in the 1930s, provide the foundation of the quantum mechanical (QM) description of particle physics and quantum field theory. The essential elements in Wigner's construction are one-particle states --- representations of the Poincar\'e group associated with a single asymptotic particle, in an irreducible representation of its little group (LG) \cite{Weinberg:1995mt}. While this satisfying picture provides the full classification of one-particle states, the general construction of multi-particle states has rarely been addressed: they are simply assumed to be direct products of one-particle states. However, in a beautiful, under-appreciated  paper in 1972 Zwanziger~\cite{Zwanziger:1972sx} found that quantum states with both electric and magnetic charges transform in non-trivial multi-particle representations of the Poincar\'e group. In the first part of this paper we address the general construction of multi-particle states and introduce the concept of the pairwise LG, which is necessary to fully classify the multi-particle representations of the Poincar\'e group. In addition to the one-particle LGs introduced by Wigner, the pairwise LG completes the characterization of the transformation properties of the multi-particle system as a whole. In particular, it may yield an additional phase under Lorentz transformations on top of the one-particle LG transformations, as in the first specific realization found by Zwanziger~\cite{Zwanziger:1972sx}. The pairwise LG is always just a $U(1)$, and in the most commonly considered scattering processes the corresponding helicity $q_{12}$ simply vanishes, confirming the expectation that the asymptotic multi-particle state is simply a direct product of the one-particle states. However, for charge-monopole scattering the  pairwise $U(1)$ helicity is the quantized ``cross product'' of charges
\begin{equation}
q_{12} = e_1 g_2 -e_2 g_1~,
\end{equation}
where $e_{1,2}$ ($g_{1,2}$) are the electric (magnetic) charges of the two particles. This implies modified transformation properties for scattering amplitudes involving both electrically and magnetically charged particles. We note that three-particle and higher LGs are always trivial, and so the general classification of multi-particle states in 4D will be given in terms of the momenta, spins/helicities and \textit{pairwise} LG helicities
\begin{equation}
\ket{ p_1,\, \ldots ,\, p_n~;~\sigma_1, \,\ldots ,\, \sigma_n~;~q_{12},\,q_{13},\, \ldots , \,q_{n-1,n} }\ .
\end{equation}

In the second half of our paper we use our refined understanding of the pairwise LG to construct scattering amplitudes of electrically and magnetically charged states. Understanding the interactions of magnetically charged states has been a long standing issue in particle physics. Dirac showed that a Lorentz invariant Lagrangian with both electric and magnetic charges must be non-local \cite{Dirac:1948um}, and such interactions are often referred to as being ``mutually non-local.'' Alternatively, Zwanziger showed \cite{Zwanziger:1970hk} that one can write a local Lagrangian, but manifest Lorentz invariance is lost. These problems seem to be an artifact of the unphysical, gauge-variant Dirac string.  For some time it was not even clear that the scattering of electrically and magnetically charged particles makes sense. Paradoxically, Weinberg found \cite{Weinberg:1965rz} that the amplitude for one photon exchange between an electric charge and a magnetic monopole  is not Lorentz invariant (and implicitly not gauge invariant \cite{Terning:2018udc}).  However, recently it was shown by Terning and Verhaaren \cite{Terning:2018udc} that an all orders resummation of soft photons can restore both Lorentz and gauge invariance if Dirac charge quantization \cite{Dirac:1931kp}  is satisfied. Hence it is believed that the electric-magnetic $S$-matrix is both local and Lorentz invariant, but Lagrangian formulations cannot make both properties manifest at the same time, leading, unsurprisingly, to seemingly unending difficulties in calculating scattering amplitudes \cite{Laperashvili:1999pu,Gamberg:1999hq,Bruemmer:2009ky,Csaki:2010rv,Sanchez:2011mf,Colwell:2015wna,Hook:2017vyc,Terning:2018lsv}. 

Thus we can see that electric-magnetic scattering is an ideal proving ground for on-shell methods. In this paper we indeed find that electric-magnetic scattering demonstrates a success for the on-shell program in theories where Lagrangian methods fall short. We should note that in our formulation we never need to introduce a Dirac string. This is in contrast to previous attempts to apply on-shell methods to electric-magnetic scattering \cite{Caron-Huot:2018ape,Huang:2019cja,Moynihan:2020gxj} which have been only partially successful in eliminating the unphysical Dirac string, thus suffering from a Lorentz violating sign ambiguity.

The bulk of our paper is devoted to extending on-shell amplitude methods to calculations of electric-magnetic $S$-matrix elements while maintaining manifest Lorentz invariance and locality. Thus we see that ``mutually non-local'' scattering is, in fact,  local aside from the angular momentum carried in the Coulomb fields of the particles.
The key is to ensure that the full action of the Poincar\'e group, including the one-particle and pairwise LGs, is properly incorporated. We find a beautiful and simple implementation of this scheme in the spinor-helicity framework, allowing us to go far beyond Zwanziger's special case of pairwise helicities equal to one. To capture the effect of the pairwise LG, we define {\it null} ``pairwise'' momenta $p_{ij}^{\flat \pm}$ which are linear combinations of the momenta of each electric-magnetic pair. The pairwise momenta are then naturally expressed using pairwise spinor-helicity variables 
\begin{equation}
\ket{p_{ij}^{\flat \pm} } , \sbra{p_{ij}^{\flat \pm}},
\end{equation}
which are constructed such that under Lorentz transformations they pick up exactly the phase dictated by the pairwise LG. Along with the standard massless and massive spinor-helicity variables, the pairwise spinor-helicity variables serve as the fundamental building blocks for the construction of the $S$-matrix for magnetic scattering\footnote{Note that we will use the term magnetic scattering or magnetic $S$-matrix to emphasize that there is at least one magnetically charged object among the scattered states, but our discussion is fully general and applicable to generic multi-dyon scattering.}.

We utilize our newly defined pairwise spinors to construct all 3-point electric-magnetic amplitudes, as a direct generalization of Arkani-Hamed, Huang, and Huang~\cite{Arkani-Hamed:2017jhn}; our derivation implies a non-trivial generalization of the selection rules derived in \cite{Arkani-Hamed:2017jhn}. For example in the decay of a massive spin $s$ to two massless particles, we get the selection rule $|\Delta h - q|\leq s$, which reduces to the standard $|\Delta h|\leq s$ in the non-magnetic case with $q=0$. Another non-trivial selection rule we derive is for the decay of a massive spin $s_1$ into two massive particles with spins $s_2$ and $s_3$. In this case we get $s_1+s_2+s_3 \geq |q_{23}|$, indicating, as a special case, that a scalar dyon cannot decay into two other scalar dyons with $q_{23}\neq0$.
Armed with our general classification of 3-point magnetic amplitudes, we move on to address the $2\to 2$ scattering of a fermion and a monopole, making use of the fully relativistic partial wave decomposition, adapted to the magnetic case. Using minimal dynamical information about the phase shifts of the higher partial wave amplitudes, we are able to fully reproduce the results of the non-relativistic quantum mechanics (NRQM) calculation of Kazama, Yang and Goldhaber~\cite{Kazama:1976fm}. In particular, our selection rules immediately tell us that in the lowest partial wave only the helicity-flip amplitudes are non-zero while forward scattering is not allowed. Furthermore, we are able to determine the full expression for the helicity flip amplitude. For the higher partial waves our formalism allows us to fix the full angular dependence of the amplitudes, while the overall magnitude of all partial waves can be fixed using unitarity and the phase shifts. 

The paper is organized as follows. Section~\ref{sec:pairwiseLG} contains our discussion of the general transformation properties of multi-particle states under the Poincar\'e group. We introduce the concept of pairwise LG here. We also give a basic introduction into the unusual properties of the charge-monopole system, rooted in the asymptotic angular momentum contained in the electromagnetic field. In section~\ref{sec:sh_monopole-fermion_$S$-matrix} we define our main objects of interest --- the pairwise spinor-helicity variables which transform covariantly under the pairwise LG. These new spinor-helicity variables, together with the standard spinors for massless and massive particles, serve as a complete set of building blocks for the magnetic (and non-magnetic) $S$-matrix. We put our new building blocks to use in section~\ref{sec:rules}, in which we demonstrate how to construct the magnetic $S$-matrix and derive concrete expressions for all magnetic 3-point amplitudes in the spirit of ref.~\cite{Arkani-Hamed:2017jhn}. In section~\ref{sec:twotwo} we take a further step and derive the general partial wave expansion for magnetic $2\rightarrow 2$ matrix elements. Finally, in sections~\ref{sec:kaza}-\ref{kazamahigher}, we apply our formalism to the case of fermion-monopole scattering, effortlessly reproducing the non-trivial results of Kazama,  Yang, and Goldhaber~\cite{Kazama:1976fm}, including the helicity-flip of the lowest partial wave and the full angular dependence of the higher partial waves.
Finally, in section \ref{sec:unit} we discuss partial wave unitarity in the context of the magnetic $S$-matrix, knowledge of which is required to obtain the magnitude of higher partial wave processes.

\section{Representations of the Poincar\'e Group for Charge-Monopole System: Pairwise LG\label{sec:pairwiseLG}}

It has long been known that the simultaneous presence of a magnetic monopole and an electric charge results in unusual rotational properties. The first explicit statement of this came from J.J. Thomson \cite{Thomson} who found that the EM field of a system containing an 
electric charge $e$ and magnetic charge $g$ carries an angular momentum even when both charges are at rest:\footnote{Due to the appearance of $E$ and $B$ the field angular momentum must be proportional to $eg$. It is also a dimensionless vector for which the only candidate is $\hat{r}$, hence the result must be proportional to $eg \hat{r}$ which can be verified by explicit calculation  \cite{Thomson}.}
\begin{equation}\label{eq:chmon}
\vec{J}^{~\text{field}}~= \frac{1}{4\pi}\,\int d^3 x ~\vec{x} \times \left( \vec{E} \times \vec{B} \right)=\,-eg\,\hat{r}~\,\equiv\,-q \hat{r}\,
\end{equation}
where $\hat{r}$ is a unit vector pointing from the magnetic monopole to the charge. Quantum mechanically, angular momentum is quantized in half integer units, and so we get yet another derivation of the Dirac quantization condition~\cite{Dirac:1931kp} $eg~=n/2$. 

The angular momentum of the electromagnetic field Eq.~(\ref{eq:chmon}) was generalized to the case of dyons by Schwinger \cite{Schwinger:1969ib} and Zwanziger \cite{Zwanziger:1969by} 
\begin{equation}\label{eq:zsh}
\vec{J}^{~\text{field}}~=~\sum\,q_{ij}\,\hat{r}_{ij}\,
\end{equation}
with the sum taken over all dyon pairs
and
\begin{equation}\label{eq:ZS}
q_{ij}~=~e_i\,g_j\,-\,e_j\,g_i~=~\frac{n}{2}\,,
\end{equation}
where the Dirac-Schwinger-Zwanziger quantization condition\footnote{Sometimes this condition is given as $(e_i\,g_j\,-\,e_j\,g_i)/4\pi~=~\frac{n}{2}$. Here and throughout we normalize the magnetic charge such that Eq.~(\ref{eq:ZS}) holds, and there is never a $(4\pi)^{-1}$ factor in the quantization condition.} for  $q_{ij}$ is once again implied by angular momentum quantization.

Zwanziger~\cite{Zwanziger:1972sx} further showed how to write the angular momentum for scattering dyons in a Lorentz covariant fashion
\begin{eqnarray}\label{eq:relZwa}
M^{\nu \rho}_{\text {field};\,\pm}~&=&~\pm\,\sum_{i>j}\,q_{i j} \,\frac{\epsilon^{\nu \rho \alpha \beta}\, p_{i \alpha}\, p_{j \beta}}{\sqrt{\left(p_{i} \cdot p_{j}\right)^{2}-m^2_{i} \,m^2_{j}}}\,,
\end{eqnarray}
where the sum is taken over all distinct dyon pairs in the initial state (final state) with a $+ (-)$ sign. The origin of the unusual $\pm$ sign is the appearance of a $t/|t|$ in the asymptotic expression for $M$.  In the non-relativistic limit, this expression reduces to $\vec{J}^{~\text{field}}_\pm=\pm\sum \,q_{ij}\,\hat{p}_{ij}$, where $\hat{p}_{ij}$ is the relative 3-momentum between the dyons in each pair. Since asymptotically  $\hat{p}\cdot\hat{r}=\mp 1$, this exactly reproduces Eq.~(\ref{eq:zsh}).

The physical implications of (\ref{eq:chmon})-(\ref{eq:zsh}) are hard to overstate. They imply the following unusual properties of charge-monopole (or general dyonic) systems: 
\begin{itemize}
\item The conserved angular momentum for the interacting theory is different from the angular momentum of the free theory  
\item As a consequence, the asymptotic quantum states representing dyon pairs do not completely factorize into single-particle states
\item In general there is no crossing symmetry   for the electric-magnetic $S$-matrix
\end{itemize} 
The first and second points can be immediately understood. Since the angular momentum of the EM field depends only on $q_{ij}$ and does not depend on the relative distance (just orientation) this term does not vanish no matter how far the charge and the monopole are separated, hence the direct product of two single-particle states never captures this additional contribution to the angular momentum. The third point will be elaborated below once we consider the LG transformation of the magnetic $S$-matrix.

\subsection{Electric-Magnetic angular momentum: the NRQM case}
 
Before jumping into our main topic, which is the representation of the Poincar{\'e} group and quantization of theories with magnetic charges, let us briefly remark on the NRQM case. Rather than defining the non-relativistic $S$-matrix in full generality, we show here how the conserved angular momentum operator $\vec{L}$ is modified in the presence of magnetic charges \cite{Lipkin:1969ck}.

The Hamiltonian of a charged particle in the background field of a stationary monopole is given by 
\bea
H = - \frac{1}{2m} \left( \vec{\nabla} - i e \vec{A} \right)^2 + V(r) = - \frac{1}{2m} \vec{D}^2 + V(r)
\eea 
where $\vec{D} = \vec{\nabla} - i e \vec{A}$ and $\vec{A}$ is the vector potential for the monopole, defined most conveniently using two coordinate-patches in \cite{Wu:1976ge}. Specifically, with the monopole at the origin, 
$A_\phi= \frac{ \pm g}{r \sin \theta} \left( 1 \mp \cos \theta \right)$ on each of the patches, usually chosen to be the upper (lower) hemisphere in the monopole rest frame. One can easily check that the usual particle definition of the angular momentum $\vec{L} = - i \vec{r} \times \vec{D}$ does not satisfy the angular momentum algebra
 \bea
&& [ L_i, L_j ] = i \epsilon_{ijk} L_k \\
&& [ L_i, H ] = 0\,.
\eea
This algebra, however, is satisfied once the angular momentum operator is generalized to include a term that depends both on electric and magnetic charges
\bea
\vec{L} = -i \vec{r} \times \vec{D} - eg \hat{r}=m \vec{r} \times \dot{\vec{r}} - eg \hat{r}\,
\label{eq:magneticL}
\eea
where $\hat{r} = \vec{r}/r$ is a unit vector pointing radially outward and we used the Heisenberg equation of motion $\dot{\vec{r}} = -i \vec{D}/m$ in the second equality. Hence for a charged particle moving in a monopole background,  angular momentum must be supplemented with an additional term proportional to $q$ corresponding to the contribution of the EM field. Importantly, the contribution of the EM field, as well as the total angular momentum, is non-vanishing even when $\dot{\vec{r}} = 0$ (i.e.~in a situation where both the charged particle and the monopole are at rest).

This expression can be generalized to a quantum field theory in the the case of a  `t Hooft-Polyakov monopole background. The 't Hooft-Polyakov monopole solution in an $SU(2)$ gauge theory is not invariant either under spatial rotations or gauge transformations, however, it is invariant under a combined transformation generated by $\vec{L} + \frac{\vec{\tau}}{2}$ (recall that the solution for the scalar field is  $\Phi_{\rm cl} \propto \tau^a {\hat r}^a$).
For a particle of spin $S$ in a representation $R$ of $SU(2)$ and moving in the monopole background, the conserved angular momentum is given by
\bea
\vec{J} = \vec{L}  + \vec{T}_R+ \vec{S}\,,
\eea
where $\vec{T}_R$ are the $SU(2)$ generators in the representation $R$. 
This expression is especially instructive  for a particle in a doublet representation of the $SU(2)$ (so that the electric charges under the unbroken $U(1)$ are minimal). In the singular gauge where the magnetic field of the monopole points in the $\tau_3$ direction in group space and the field contribution to the angular momentum is $\pm 1/2$, we find an exact match to the NRQM result. In the relativistic quantum theory, this extra contribution gives rise to the additional LG phase, as we discuss below.

\subsection{Pairwise LG\label{sec:pairwiseLGsub}}

In order to properly understand the effect of the modified angular momentum operator on the construction of the quantum mechanical Hilbert space we first need to go back and understand the properties of multi-particle representations of the Poincar{\'e} group. It is well-known that for single particles one needs to define a reference momentum $k$, which may be chosen as $(M,0,0,0)$ for massive particles or $(E ,0,0,E )$ for massless particles. The LG is then the set of Lorentz transformations that leave the reference momenta invariant. For massive particles the LG is $SO(3) \sim SU(2)$, while for massless particles it is $ISO(2)$ the two dimensional Euclidean group. The nature of the particle we are describing thus determines the required  representation of the LG. For example, given a massive particle the representation is specified by the mass and the spin, $s$, and the state in the Hilbert space is just $| k,s \rangle$. For the case of massless particles, while interesting non-trivial representations of $ISO(2)$ are in principle allowed by the kinematics of the Lorentz group~\cite{Schuster:2014hca}, the models needed to match experiment do not take advantage of the additional quantum number offered by using the entire $ISO(2)$ group rather than just the $SO(2) \sim U(1)$ subgroup corresponding to ordinary helicity. 

When considering the representations of the Poincar{\'e} group one usually stops here and assumes that multi-particle states transform  as products of single particle states. However a closer examination of the  Poincar{\'e}  group shows that this is not the only possibility: as first pointed out by Zwanziger~\cite{Zwanziger:1972sx}, there are rotations that leave the momenta of a pair of particles invariant. To see this, we can consider a two-particle state $\vert p_1, p_2 \rangle$ and again consider some reference momenta for this multi-particle state. The simplest choice is to go into the center of momentum (COM) frame
\begin{eqnarray}\label{eq:refmom}
&& (k_1)_{\mu}~=~\left(E^c_1,0,0,+\,p_c\right)\,\nonumber\\
&& (k_2)_{\mu}~=~\left(E^c_2,0,0,-\,p_c\right)\, ,\label{eq:k_i_and_k_j}
 \end{eqnarray}
where
\begin{eqnarray}\label{eq:ps}
p_c~=~\sqrt{\frac{(p_1 \cdot p_2)^2 - m_1^2 m_2^2}{s}}~~,~~E^c_{1,2}=\sqrt{m^2_{1,2}+p^2_{\text{c}}}\,,
 \end{eqnarray}
are Lorentz invariant, and $s=(E^c_1+E^c_2)^2$. A Lorentz boost $L_p$ brings the reference momenta back into the arbitrary pair of original momenta $p_1= L_p \,k_1,~p_2 = L_p \,k_2$. The important observation is that there exists a non-trivial two-particle or pairwise LG which leaves these reference momenta unchanged --- it is simply a rotation around the $z$-axis, corresponding to a $U(1)$ pairwise LG. We would like to emphasize that this pairwise LG is independent of the usual one-particle LG: it describes the relative transformation of the two particle state 
compared to the product of the one-particle states. Hence the general two-particle state is characterized by the representations of the individual particles under the one-particle LG, as well as the additional $U(1)$ charge, $q_{12}$, corresponding to the representation of the two-particle state under the pairwise LG. We call this charge the ``pairwise helicity''.
Thus the state is  $\ket{\,p_1,p_2~;~\sigma_1,\sigma_2~;~q_{12}\,}$. The $p_1,\,p_2$ are simply the individual momenta for each particle, and the $\sigma_i$ are collective indices denoting the individual $s^2_i
,\,s^z_i$ for massive particles or the helicity $h_i$ for massless ones. The novelty here is the additional quantum number $q_{12}$, which is associated with the particle \textit{pair} rather than an individual particle. Under a Lorentz transformation, this quantum state transforms as 
\begin{equation}
U(\Lambda )~\ket{\,p_1,p_2~;~\sigma_1,\sigma_2~;~q_{12}\,}~=~e^{i \,q_{12} \,\phi} ~\mathcal{D}_{\sigma'_1\,\sigma_1}\,\,\mathcal{D}_{\sigma'_2\,\sigma_2}~\ket{\,\Lambda p_1,\Lambda p_2~;~\sigma'_1,\sigma'_2~;~q_{12}\,}
\end{equation}
where $\phi$ is the $U(1)$ rotation angle corresponding to the pairwise LG, while the $\mathcal{D}$s' are representations of one-particle LG rotations for each of the two particles. For massive particles, the LG is just $SU(2)$ and the $\mathcal{D}$ matrices are in the spin $s_i$ representation of $SU(2)$. For massless particles, the LG is $U(1)$ and the $\mathcal{D}$s are the ordinary helicity phases $e^{ih_i\phi_i}$. We will show that this is indeed the right transformation for the spinless case, and leave the general case for future work.

This transformation rule can be derived in the usual way by applying Wigner's method of induced representations~\cite{Wigner:1939cj} , which we briefly summarize at the end of this subsection. But first we would like to ask  what happens for the case of more than two particles. To that end it is sufficient to consider a three particle state. Clearly, its transformation includes a product of three representations of the one-particle LG. Each one-particle LG transformation leaves the momentum of the corresponding particle invariant. The three particle state also transforms as a product of representations under three pairwise LGs, each leaving the momenta of the corresponding pair invariant. 
However, there is no non-trivial subgroup of the Poincar{\'e}  group that leaves invariant an  arbitrary set of  three momenta. Hence the three-particle LG is trivial and the Lorentz transformations of three particle states are fully characterized by their transformations under three single particle LGs and three pairwise LGs. This conclusion easily generalizes to all $n$-particle states: such states are characterized by $n$ masses and spins, as well as ${n\choose 2}$ pairwise $U(1)$ helicities $q_{ij}$, 
$\ket{\,p_1,p_2,\ldots, p_n~;~\sigma_1,\sigma_2, \ldots ,\sigma_n~;~q_{12}, q_{13},\ldots ,q_{n-1,n}\,}$ with Lorentz transformations given by 
\begin{eqnarray}
& U(\Lambda )\,\ket{\,p_1,\ldots ,p_n~;~\sigma_1,\ldots ,\sigma_n~;~ q_{12}, q_{13},\ldots q_{n-1,n}\,} = &
\nonumber \\
& e^{i\sum_{i<j}  q_{ij} \phi(p_i,p_j,\Lambda ) }~\prod_{i=1}^n \,\mathcal{D}^{i}_{\sigma'_i\sigma_i}  \ket{\,\Lambda p_1,\ldots ,\Lambda p_n~;~\sigma_1',\ldots , \sigma_n'~;~q_{12}, q_{13}, \ldots , q_{n-1,n}\,} &\ .\nonumber\\
\label{eq:generalLorentz}
\end{eqnarray}
The exact representations of the pairwise LGs for multi-particle states, i.e. the helicities $q_{ij}$, depend on the dynamics of the theory. In most cases only trivial representations of the pairwise LGs arise and $q_{ij}=0$. The one known exception is a state containing both electric and magnetic charges. As we will see below, the action of the angular momentum operator requires in this case the identification $q_{ij} =  e_i g_j -e_j g_i$, corresponding to the Dirac-Schwinger-Zwanziger quantization condition; the existence of EM field angular momentum implies that multi-particle states do not fully factorize into products of single particle states.

We conclude this subsection by reviewing the Wigner method of induced representations to derive Eq.~(\ref{eq:generalLorentz}) for the spinless case with two particles, following~\cite{Wigner:1939cj,Weinberg:1995mt,Zwanziger:1972sx}. This also provides us with an explicit formula for the pairwise LG phase $\phi (p_i,p_j, \Lambda)$.  We define our reference quantum states as 
\begin{eqnarray}
\ket{\,k_1, k_2~;~q_{12}\,}\, .
\end{eqnarray}
Having identified  the effect of the pairwise LG on the reference states with a rotation around $z$-axis we have
 \begin{eqnarray}\label{eq:eigen}
J_z~\ket{\,k_1, k_2~;~q_{12}\,} = q_{12}\,\ket{\,k_1, k_2~;~q_{12}\,}\, .
\end{eqnarray}
This equality correctly reproduces the EM field contribution to the angular momentum in Eqs.~(\ref{eq:zsh})-(\ref{eq:relZwa}) provided that  $q_{ij}=e_ig_j-e_jg_i$. 
Interestingly, this identification also directly implies the Dirac-Schwinger-Zwanziger condition for $q_{12}$, simply from the properties of the Lorentz group. To see this, note that due to the spinorial double coverings of the Lorentz group, any $4\pi$ rotation (rather than $2\pi$) around $\hat{z}$  must be the identity,
\bea
e^{i 4\pi q_{12}} = 1 \;\; \Rightarrow \;\; q_{12}~\equiv~e_1\,g_2 \,-\, e_2\,g_1~=~\frac{n}{2}, \; n \in \mathbb{Z}.
\eea

The quantum states for general momenta $p_1,\,p_2$ can be obtained from the reference pairwise state with a Lorentz boost
\bea\label{eq:canonical}
\ket{\,p_1, p_2~;~q_{12}\,} \equiv U \left( L_p \right) \ket{\,k_1, k_2~;~q_{12}\,}\, ,
\eea
where $U(L_p)$ is a unitary operator representing the Lorentz boost $L_p$.
We now wish to learn how a generic Lorentz transformation $\Lambda$ acts on the states $\ket{\,p_1,  p_2\,;\,q_{12}\,}$. Proceeding as in the standard method of induced representations, we have
\bea
U (\Lambda) \ket{\,p_1, p_2~;~q_{12}\,} &=& U \left( L_{\Lambda p} \right) \,U \left(  L_{\Lambda p}^{-1} \Lambda L_p \right) \ket{\,k_1, k_2~;~q_{12} \,}  \nonumber \\
&=& U \left( L_{\Lambda p} \right)\, U \left( W_{k_1,k_2} \right)\ket{\,k_1, k_2~;~q_{12}\, } ,
\eea
where $W_{k_1,k_2} (p_1,p_2,\Lambda)\equiv L_{\Lambda p}^{-1} \Lambda L_p=R_z\left[\phi (p_1,p_2,\Lambda)\right]$ is a LG transformation, which is nothing but a rotation around the $z$-axis with an angle $\phi (p_1,p_2,\Lambda)$. By definition, this LG transformation  acts on $\ket{k_1, k_2~;~q_{12}}$ as $\exp\left[i q_{12} \phi (p_1,p_2,\Lambda) \right]$, so that
\bea\label{eq:twophase}
U (\Lambda) \ket{\,p_1, p_2~;~q_{12}\,} = e^{ i q_{12} \phi (p_1,p_2,\Lambda)}~ \ket{\, \Lambda p_1, \Lambda p_2~;~q_{12}\,}.
\label{eq:pole-charge_Lorentz_transf_spinless}
\eea

 We can easily see that the transformation rule for general multi-particle states in Eq.~(\ref{eq:generalLorentz})  is unitary and indeed forms a representation of the Lorentz group.  First, since Eq.~(\ref{eq:generalLorentz}) only differs from the standard Lorentz transformation by a phase $e^{i\Sigma}$, this transformation is clearly unitary. Second, because the phase angles $\phi(p_i,p_j,\Lambda)$ are identical to the ones that arose as LG phases for the two-scalar case, and since they furnish a representation, 
we know that  %
\bea
\phi(p_i, p_j,\Lambda_2\Lambda_1)~=~\phi(\Lambda_1 p_i, \Lambda_1 p_j,\Lambda_2)+\phi(p_i, p_j,\Lambda_1)\, .
\eea
This proves that $U(\Lambda_2\Lambda_1)=U(\Lambda_2)\,U(\Lambda_1)$ and so our transformation rule is indeed a representation of the Lorentz group.

\subsection{In- and Out-states for the Electric-Magnetic $S$-matrix}\label{subsec:monopole-charge_asymp_state}

Now that we understand the general transformation properties of dyonic multi-particle states, we are ready to define the relativistic $S$-matrix for electric-magnetic scattering processes. To do this we have to first properly define the multi-particle \textit{in}- and \textit{out}- states.
As usual, we separate the full Hamiltonian of the system into a free Hamiltonian, $H_0$, and an interaction:
 \begin{eqnarray}\label{eq:Hami}
H~=~H_0\,+\,V\,.
 \end{eqnarray}
In the standard definition, due to Weinberg \cite{Weinberg:1995mt}, we can choose our quantum in/out states to be eigenstates of the full interacting Hamiltonian that approach free states\footnote{Actually this language is not completely accurate since the in/out- states are conventionally defined in the Heisenberg picture and are time independent. For a rigorous definition of our $S$-matrix, see appendix~\ref{sec:Smat}.} as $t\rightarrow\pm\infty$. However, in the case of electric-magnetic scattering, this definition has to be modified. This is because $H_0$ and  $H$ have \textit{different} conserved angular momentum operators,
\begin{eqnarray}\label{eq:angmomops1}
\bks{H,\vec{J}}~=~\bks{H_0,\vec{J}_{0}}~=~0,~~~\vec{J}~\neq~\vec{J}_{0}\, .
\end{eqnarray}
The operator $J_0$ represents the total orbital and spin angular momentum of different particles, while $J$ also includes the contribution of the EM field, as is evident from Eq.~(\ref{eq:relZwa}). The inequality of $J$ and $J_0$  seems, so far,  to be unique to electric-magnetic scattering. As a consequence the Lorentz group is represented differently\footnote{The generator of boosts $K$ is always represented on the \textit{in}/\textit{out} states differently from its representation on free states. The surprise here is the difference between \textit{in}- and \textit{out}- states, which is a unique consequence of $J\neq J_0$.} on the \textit{in}- and \textit{out}- eigenstates of $H$. This is simply a reflection of the fact that $q_{ij}$ can be non-vanishing for the \textit{in}- and \textit{out} states, while the eigenstates of $H_0$ are simply the direct product states of the free one-particle states with all $q_{ij}=0$.

In accordance with our discussion in section~\ref{sec:pairwiseLGsub}, we identify the multi-particle \textit{in}- and \textit{out}-states as the states transforming with definite values of $q_{ij}$:  
\bea\label{eq:2momtrans}
U(\Lambda)\, \ket{p_1, \ldots ,p_n\,;\,\pm\,}~&=&~\prod_i \mathcal{D} (W_i)~\ket{\Lambda p_1, \ldots ,\Lambda p_n\,;\,\pm\,}\,e^{\pm i\,\Sigma}\, ,
\eea
where $\Sigma\equiv\sum_{i>j}^n q_{ij}\, \phi (p_i,p_j, \Lambda)$. Here, and below, `$+$' stands for `in', and `$-$' stands for `out', the $\mathcal{D} (W_i)$ are the one-particle LG transformations,  while the  $e^{\pm i\, \Sigma}$ is the additional phase factor corresponding to the pairwise LGs. Note that we need to choose opposite signs for the pairwise LG phases for the  \textit{in}- and \textit{out}- sates, in accordance with the extra sign showing up in the asymptotic expression (\ref{eq:relZwa}). 
We see that the transformation rule Eq.~(\ref{eq:2momtrans}) is a departure from Weinberg's standard definition of the $S$-matrix, in the sense that the Lorentz group is represented \textit{differently} on \textit{in}- and \textit{out}- states.

\subsection{Lorentz transformation of the electric-magnetic $S$-matrix}
\label{subsec:monopole-charge_S_matrix:Lorentz_transf}

In the previous section, we presented the Lorentz transformation, Eq.~(\ref{eq:2momtrans}), of multi-particle quantum states involving electric and magnetic charges. The general LG transformation for the $S$-matrix readily follows,
\bea\label{eq:Lorentz_transf_$S$-matrix_charge_pole_1} 
&& S \left(p'_1,\ldots,p'_m\,|\, p_1,\ldots,p_n\right)~\equiv~\bk{p'_1,\ldots,p'_m ;\, -\,| \, p_1,\ldots,p_n;\, +\,} \nonumber \\
&&=~\bk{p'_1,\ldots,p'_m ;\, -\,|  U (\Lambda)^{\dagger} ~U(\Lambda) |\, p_1,\ldots,p_n;\, +\,} \nonumber\\
&&=~e^{i(\Sigma_++\Sigma_-)}~\prod_{i=1}^m \mathcal{D} (W_{i})^{\dagger}\,~\prod_{j=1}^n \mathcal{D}(W_{j}),~S \left(\Lambda\,p'_1,\ldots,\Lambda\,p'_m\,| \,\Lambda\,p_1,\ldots,\Lambda\,p_n\right)
\eea
where\,\footnote{Below we use the notation $\phi_{ij}=\phi (p_i,p_j, \Lambda)$ when it's clear whether we are talking about the \textit{in}- or \textit{out}- state.} 
\bea
\Sigma_+\equiv\sum_{i>j}^n q_{ij}\, \phi (p_i,p_j, \Lambda)~~~~,~~~~\Sigma_-\equiv\sum_{i>j}^m q_{ij} \,\phi (p'_i,p'_j, \Lambda)\,.
\eea
 and $W_{i}$ are the LG rotations for one-particle states in the \textit{in}- and \textit{out}- states. To go from the second to the third line, we used the fact that the extra $U(1)$ LG factor has the same sign for $\bra{out}$ and $\ket{in}$ states. Note that since $\Sigma_\pm$ pairs particles within the \textit{in}- and \textit{out}- states but doesn't involve in-out pairs, this is a manifest violation of crossing symmetry. Inverting Eq.~(\ref{eq:Lorentz_transf_$S$-matrix_charge_pole_1}), we have
\bea\label{eq:Lorentz_transf_$S$-matrix_charge_pole_2} 
&& S \left(\Lambda\,p'_1,\ldots,\Lambda\,p'_m\,|\, \Lambda\,p_1,\ldots,\Lambda\,p_n\right)~=~\nonumber\\
&&e^{-i\,(\Sigma_++\Sigma_-)}~\prod_{i=1}^m \mathcal{D} (W_{i})~\prod_{j=1}^n \mathcal{D}(W_{j})^{\dagger}\,~S\left(p'_1,\ldots,p'_m\,| \,p_1,\ldots,p_n\right)
\eea
This transformation rule was first derived in \cite{Zwanziger:1972sx}. If all $q_{ij}=0$ (in particular, if none of the scattering particles have magnetic charge), the transformation rule Eq.~(\ref{eq:Lorentz_transf_$S$-matrix_charge_pole_2}) reduces to the standard LG transformation with $\Sigma_\pm=0$.
To construct the electric-magnetic $S$-matrix elements  that satisfy the transformation rule given in Eq.~(\ref{eq:Lorentz_transf_$S$-matrix_charge_pole_2}) using on-shell methods we need to introduce a new kind of spinor-helicity variable that enables us to saturate the extra ``electric-magnetic'' $U(1)$ phase in Eq.~(\ref{eq:Lorentz_transf_$S$-matrix_charge_pole_2}).

\section{Pairwise Spinor-Helicity Variables for the Electric-Magnetic $S$-matrix}
\label{sec:sh_monopole-fermion_$S$-matrix}

\label{subsec:flat_variable}

\subsection{Standard spinor-helicity variables for the standard LG}
In the spinor-helicity formalism without magnetic charges, we can directly write down the amplitude that transforms by construction as in Eq.~(\ref{eq:Lorentz_transf_$S$-matrix_charge_pole_2}) with $q=0$. To do this, we construct the amplitude from contractions of the spinor-helicity variables. For a massless particle $i$, we use the spinor-helicity variables $\ket{p_i}_{\alpha}\,\sbra{p_i}_{\dot{\alpha}}$, which transform under Lorentz transformations as
  \begin{eqnarray}\label{eq:transfstand}
\Lambda^{~\beta}_{\alpha}\ket{p_i}_{\beta}~&=&~ e^{+ \frac{i}{2}\phi(p_i,\Lambda)}\,\,\ket{\Lambda p_i}_{\beta} , \hspace{0.5cm} 
\sbra{p_i}_{\dot{\beta}}\,\tilde{\Lambda}^{\dot{\beta}}_{~\dot{\alpha}} ~=~e^{- \frac{i}{2}\phi(p_i,\Lambda)}\,\,\sbra{\Lambda p_i}_{\dot{\alpha}}\,,
 \end{eqnarray}
where the phase $\phi (p_i,\Lambda )$ corresponds to the action of the one-particle LG for massless particles. For a derivation of this transformation rule, see for example \cite{Elvang:2013cua,Henn:2014yza,Cheung:2017pzi}. In many cases we simply drop the $p_i$ from the spinors and just use the notation $\ket{i}_{\alpha}\equiv\ket{p_i}_{\alpha}$ and $\sbra{i}_{\dot{\alpha}}\equiv\sbra{p_i}_{\dot{\alpha}}$. An $S$-matrix involving an outgoing massless particle $i$ with helicity $h_i$ has the correct LG phase for the $i^{th}$ particle if we construct it from $n_i$ copies of $\ket{i}_{\alpha}$ and $\tilde{n}_i$ copies of $\sbra{i}_{\dot{\alpha}}$, such that $\tilde{n}_i-n_i\,=\,2\,h_i$.\footnote{Notice that while $\vert p \rangle$ ($\vert p ]$) carries a helicity weight $\pm 1/2$, as is evident from Eq.~(\ref{eq:Lorentz_transf_$S$-matrix_charge_pole_2}), for checking LG scaling of the $S$-matrix, we need to do $\vert p \rangle \to \vert \Lambda p \rangle \propto \omega^{-1} \vert p \rangle$ and $\vert p ] \to \vert \Lambda p ] \propto \omega \vert p ]$, where $\omega$ is a helicity $+1/2$ factor. }

Similarly, an amplitude involving a massive particle $j$ of spin $s_j$ is constructed from $2s_j$ insertions of the \textit{massive} spinor-helicity variables $\ket{\mathbf{i}}^I_{~\alpha}$, with their spinor indices symmetrized. The indices $I$ on the massive spinors indicate that they transform as doublets of the LG $SU(2)$ for massive particles. These indices are usually suppressed, as they are only needed when taking the massless limit (specifying a value for the $I$ index is like choosing a particular helicity in the massless limit). Note that the ${I}$ indices are automatically symmetrized when one symmetrizes over the spinor indices $\alpha$ or $\dot{\alpha}$. We refer the reader to ref.~\cite{Arkani-Hamed:2017jhn} for a detailed discussion of the spinor-helicity formalism for massive particles. 
 
\subsection{Pairwise momenta}
As we argued in the previous section, in the case of the electric-magnetic $S$-matrix\footnote{In our construction for electric-magnetic scattering we refer to the  ``$S$-matrix'' rather than the usual scattering amplitude. The reason behind this is that in the magnetic case, selection rules sometimes forbid the appearance of the $\delta$ function in the standard relation $S_{\alpha\beta}=\delta(\alpha-\beta)\,-\,2i\pi\delta^{(4)}(p_\alpha-p_\beta)\,\mathcal{A}_{\alpha\beta}$.}, the transformation rule involves an additional pairwise LG phase associated with the angular momentum in the EM field, as can be seen in Eq.~(\ref{eq:Lorentz_transf_$S$-matrix_charge_pole_2}). Since this extra phase is associated with pairs of momenta $p_i,\,p_j$, it is not possible to reproduce the correct transformation rule using only the standard spinor-helicity variables $\ket{i}_{\alpha}$ and $\sbra{i}_{\dot{\alpha}}$ (or $\ket{\mathbf{i}}^I_{~\alpha}$ and $\sbra{\mathbf{i}}^I_{\dot{\alpha}}$). This motivates us to the define a new kind of spinor-helicity variable associated with pairs of momenta $p_i,\,p_j$, which transform with the pairwise LG phase $\phi_{ij}$. Importantly, the pairwise LG transformation of the $S$-matrix is always a $U(1)$ phase, and so we need the new spinors to be \textit{massless}, and associated with \textit{null} momenta.  

Since the extra LG factor for the electric-magnetic $S$-matrix is associated with the momenta $p_i,\,p_j$ of each pair in the in/out- state, it is natural to define two null linear combinations of $p_i,\,p_j$, which we call 
the pairwise momenta\footnote{The use of the label $\flat$ to denote null linear combinations of timelike momenta is inspired by the notation of \cite{Kosower:2004yz} and of the OPP reduction \cite{Ossola:2006us} in the context of generalized unitarity \cite{Bern:1994cg,Forde:2007mi}. There, null combinations of external momenta were used in order to construct a null basis to span the internal loop momenta that have been put on shell.} $p_{ij}^{\flat\pm}$. Below, we will define pairwise spinor-helicity variables associated with these pairwise momenta, and show that they have the correct pairwise LG weight to be used as building blocks for the electric-magnetic $S$-matrix. We first define the ``reference'' pairwise (null) momenta in the COM frame as
\begin{eqnarray}\label{eq:kpmeq}
\left(k^{\flat\pm}_{ij}\right)_\mu~=~p_c\left(1,0,0,\pm1\right)\, ,
\end{eqnarray}
where $p_c$ is the COM momentum of the $ij$ pair, as in Eq.~(\ref{eq:ps}). The pairwise momenta $p^{\flat\pm}_{ij}$ in any other frame can be obtained by boosting $k^{\flat\pm}_{ij}$ into that frame. Clearly $k^{\flat\pm}_{ij}\cdot k^{\flat\pm}_{ij}=0$ and $k^{\flat+}_{ij}\cdot k^{\flat-}_{ij}=2p^2_c$, and these relations obviously hold in any other frame.

For reference, we also present the Lorentz covariant definition of $p^{\flat\pm}_{ij}$,
\begin{eqnarray}\label{eq:ppmeq}
p^{\flat+}_{ij} &=& \frac{1}{E^c_i+E^c_j} \left[\left(E^c_j+p_c\right) p_i - \left(E^c_i-p_c\right)  p_j \right]\nonumber\\
p^{\flat-}_{ij} &=& \frac{1}{E^c_i+E^c_j} \left[\left(E^c_i+p_c\right) p_j - \left(E^c_j-p_c\right)  p_i \right]\, .
\end{eqnarray}

In the $m_i\rightarrow 0$ limit, we have $E^c_i\rightarrow p_c$ and so $p^{\flat+}_{ij}\rightarrow p_i$ and $p^{\flat-}_{ij}$ becomes Parity-conjugate of $p_i$. Similarly, in the $m_j\rightarrow 0$ limit, we have $E^c_j\rightarrow p_c$ and so $p^{\flat-}_{ij}\rightarrow p_j$ and $p^{\flat+}_{ij}$ becomes Parity-conjugate of $p_j$.
By inverting these equations, we can express the massive momenta using the null momenta as
\begin{eqnarray}\label{eq:ppmeqinv}
p_i &=& \frac{1}{2p_c} \left[\left(E^c_i+p_c\right) p^{\flat+}_{ij} + \left(E^c_i-p_c\right)  p^{\flat-}_{ij}\right]\nonumber\\
p_j &=& \frac{1}{2p_c} \left[\left(E^c_j+p_c\right) p^{\flat-}_{ij} + \left(E^c_j-p_c\right)  p^{\flat+}_{ij}\right]\,.
\end{eqnarray}

 \subsection{Pairwise spinor-helicity variables}\label{sec:pairwise}

We are now in a position to define spinor-helicity variables related to the pairwise  momenta $p^{\flat\pm}_{ij}$. As we will show, these pairwise spinor-helicity variables transform with a $U(1)$ LG phase directly related to the pairwise LG phase of the \textit{in}- and \textit{out}- states in Eq.~(\ref{eq:2momtrans}). This makes them natural building blocks for the electric-magnetic $S$-matrix.

As a first step, note that linearity implies that the canonical Lorentz transformation $L_p$ defined in Eq.~(\ref{eq:canonical}) that takes $k_{i}\rightarrow p_{i}$ also gives
 \begin{eqnarray}\label{eq:Lpppm}
L_p \,k^{\flat\pm}_{ij}~&=&~p^{\flat\pm}_{ij}\, .
 \end{eqnarray}
This is instrumental in proving that the pairwise spinor-helicity variables defined below transform with the same LG phase as the two-particle states in Eq.~(\ref{eq:pole-charge_Lorentz_transf_spinless}).
The next step is to define the reference pairwise spinor-helicity variables,
 \begin{eqnarray}
&&\ket{k^{\flat+}_{ij}}_{\alpha}~=~\sqrt{2\,p_c}\,\colvec{1\\ 0}~~~,~~~\,\ket{k^{\flat-}_{ij}}_{\alpha}~=~\sqrt{2\,p_c}\,\colvec{0\\1}\nonumber\\
&&\sbra{k^{\flat+}_{ij}}_{\dot{\alpha}}~=~\sqrt{2\,p_c}\,\left(1~ ~0\right)~~~,~~~~\sbra{k^{\flat-}_{ij}}_{\dot{\alpha}}~\,=~\sqrt{2\,p_c}\,\left(0~ ~1\right)\, .
 \end{eqnarray}
These spinors are the ``square roots'' of the null reference momenta
 \begin{eqnarray}
k^{\flat\pm}_{ij} \cdot \sigma_{\alpha{\dot{\alpha}}}~&=&~\ket{k^{\flat\pm}_{ij}}_{\alpha}\sbra{k^{\flat\pm}_{ij}}_{\dot{\alpha}}\, .
 \end{eqnarray}
The above relation is a standard mapping of a bi-spinor into a vector. Multiplying both sides by $\bar\sigma_\nu^{\dot\alpha\alpha}$ and taking the trace we can also write it in the form
\begin{equation}
 2 \left( k^{\flat\pm}_{ij} \right)^\nu =\bra{k^{\flat\pm}_{ij}}^{\alpha}\sigma^\nu_{\alpha\dot{\alpha}}\sket{k^{\flat\pm}_{ij}}^{\dot{\alpha}}\,.
\end{equation}
While the LHS of this relation transforms with $L_p$ under a Lorentz transformation, the helicity variables on the RHS transform with ${\left(\mathcal{L}_p\right)}^{~\beta}_{\alpha}$ and ${\left(\tilde{\mathcal{L}}_p\right)}^{\dot{\beta}}_{~\dot{\alpha}}$ appropriate for spinorial representation. Thus up to a LG invariant factor the pairwise spinors $p_{ij}^{\flat\pm}$ are defined by
 \begin{eqnarray}
&&\ket{p^{\flat\pm}_{ij}}_{\alpha}~=~ \left(\mathcal{L}_p \right)^{~\beta}_{\alpha}\,\ket{k^{\flat\pm}_{ij}}_\beta~~~,~~
\sbra{p^{\flat\pm}_{ij}}_{\dot{\alpha}}~=~\sbra{k^{\flat\pm}_{ij}}_{\dot{\beta}}\,\left( \tilde{\mathcal{L}}_p \right)^{\dot{\beta}}_{~\dot{\alpha}}\nonumber\\.
 \end{eqnarray}
This guarantees the relation
 \begin{eqnarray}
p^{\flat\pm}_{ij} \cdot \sigma_{\alpha{\dot{\alpha}}}~&=&~\ket{p^{\flat\pm}_{ij}}_{\alpha}\sbra{p^{\flat\pm}_{ij}}_{\dot{\alpha}}\, .
 \end{eqnarray}
Following the same procedure as in the standard definition of spinor-helicity variables, it is straightforward to show that they transform with a $U(1)$ LG factor as required, since
  \begin{eqnarray}\label{eq:transf}
\Lambda^{~\beta}_{\alpha}\ket{p^{\flat\pm}_{ij}}_{\beta}~&=&~ e^{\pm \frac{i}{2}\phi(p_i,p_j,\Lambda)}\,\ket{\Lambda p^{\flat\pm}_{ij}}_{\alpha}~~,~~
\sbra{p^{\flat\pm}_{ij}}_{\dot{\beta}} \,\tilde{\Lambda}^{\dot{\beta}}_{~\dot{\alpha}} ~=~e^{\mp \frac{i}{2}\phi(p_i,p_j,\Lambda)}\,\sbra{\Lambda p^{\flat\pm}_{ij}}_{\dot{\alpha}}.\nonumber\\
 \end{eqnarray}
 Where $\Lambda^{~\beta}_{\alpha}$ and $\tilde{\Lambda}^{\dot{\beta}}_{~\dot{\alpha}}$ are the spinor versions of the Lorentz transformation $\Lambda$. Note that $\ket{p^{\flat+}_{ij}}_{\alpha}$ and $\ket{p^{\flat-}_{ij}}_{\beta}$ have \textit{opposite} pairwise helicities $\pm 1/2$. Importantly, the LG phase $\phi(p_i,p_j,\Lambda)$ in Eq.~(\ref{eq:transf}) is defined with respect to the canonical Lorentz transformation $L_p$, which is the same as the one we used to derive the transformation rule of the quantum states in section~\ref{eq:pole-charge_Lorentz_transf_spinless}. This proves that $\phi(p_i,p_j,\Lambda)$ is exactly the same phase as the one in Eq.~(\ref{eq:pole-charge_Lorentz_transf_spinless}). Consequently, we are free to use our pairwise spinor-helicity variables to construct an $S$-matrix that transforms correctly under the pairwise (and also one particle) LGs.
Explicit expressions for spinor-helicity variables in the COM frame are given in appendix~\ref{sec:explicit_fermion_monopole_var}. Here we simply present the main results in the $m_i\rightarrow 0 $ limit:
\begin{eqnarray}\label{eq:COMforms}
&&\bks{p^{\flat+}_{ij}\,i}~=~\bk{i\, p^{\flat+}_{ij}}~=~\bks{\hat{\eta}_i\, p^{\flat-}_{ij}}~=~\bk{p^{\flat-}_{ij}\,\hat{\eta}_i}~=~0\nonumber\\
&&\bks{p^{\flat-}_{ij}\,i}~=~\bk{i\, p^{\flat-}_{ij}}~=~\sqrt{2p_c}\bks{\hat{\eta}_i\, p^{\flat+}_{ij}}~=~\sqrt{2p_c}\bk{ p^{\flat+}_{ij}\,\hat{\eta}_i}~=~2p_c\, ,
\end{eqnarray}
where $\ket{i }_\alpha,\,\sbra{i}_{\dot{\alpha}}$ are the standard massless spinor-helicity variables, and $\ket{\hat{\eta}_i }_\alpha,\,\sbra{\hat{\eta}_i}_{\dot{\alpha}}$ are the (dimensionless) Parity-conjugate massless spinors that appear in the massless limit of the massive spinors $\ket{\mathbf{i} }^I_\alpha,\,\sbra{\mathbf{i}}^I_{\dot{\alpha}}$ (see ref.~\cite{Arkani-Hamed:2017jhn} for their definition). Note that the above equations are Lorentz \textit{and} LG invariant, and so hold in any other reference frame as well.

\section{Constructing Electric-Magnetic $S$-matrices}
\label{sec:rules}

In section~\ref{subsec:monopole-charge_S_matrix:Lorentz_transf} we derived the transformations of electric-magnetic $S$-matrices under the pairwise and one-particle LGs:
\bea\label{eq:redef}
&& S \left(\Lambda\,p'_1,\ldots,\Lambda\,p'_m\,|\, \Lambda\,p_1,\ldots,\Lambda\,p_n\right)~=~\nonumber\\
&&e^{-i\,(\Sigma_-+\Sigma_+)}~\prod_{i=1}^m \mathcal{D} (W_{i})~\prod_{j=1}^n \mathcal{D}(W_{j})^{\dagger} \,~S \left(p'_1,\ldots,p'_m\,| \,p_1,\ldots,p_n\right)
\eea
To make use of this transformation for constructing electric-magnetic $S$-matrix elements, we defined the pairwise spinor-helicity variables in section~\ref{sec:pairwise}. Now we can use the pairwise and regular spinor-helicity variables to construct $S$-matrices that respect Eq.~(\ref{eq:redef}). This enables us to fix electric-magnetic $S$-matrix elements up to a LG invariant. 

We also reiterate here  that we are constructing electric-magnetic \textit{$S$-matrix elements} rather than \textit{amplitudes}. 
This is because by using the word ``amplitude'' we are implicitly assuming the possibility of forward scattering, as encoded in the standard relation
\begin{eqnarray}\label{eq:rel}
S_{\alpha\beta}~=~\delta(\alpha-\beta)\,-\,2i\pi\delta^{(4)}(p_\alpha-p_\beta)\,\mathcal{A}_{\alpha\beta}\, .
\end{eqnarray}
However, in our very peculiar case of electric-magnetic scattering, the decomposition of Eq.~(\ref{eq:rel}) may not actually hold. In fact, we will see below that selection rules generically forbid forward scattering for the lowest partial wave, which makes the relation Eq.~(\ref{eq:rel}) inadequate for electric-magnetic scattering. Rather than trying to adapt it to our case, we opt to never use this relation at all and just construct the $S$-matrix itself directly. Energy and momentum conservation are implicitly assumed. 

In constructing the $S$-matrix we use an all-outgoing convention common in the amplitudes literature. However, the use of this convention in the study of magnetic $S$-matrix elements is non-trivial due to lack of crossing symmetry in electric-magnetic scattering. Thus we begin by reviewing the subtleties associated with the all-outgoing convention.

\subsection{The all-outgoing convention}\label{sec:allout}

In section~\ref{subsec:monopole-charge_S_matrix:Lorentz_transf}, we described how general electric-magnetic $S$-matrices transform under Lorentz transformations. In that section, the discussion was in terms of \emph{in}- and \emph{out}-states. In the spinor-helicity formalism it is however customary to use a notation where all particles are outgoing which we call the \emph{out-out} formalism. In the standard cases without magnetic charges this is achieved using the crossing symmetry of the $S$-matrix.
To define crossing symmetry, we first assume analyticity, namely, that the $S$-matrix is an analytic function of its complexified external momenta. Crossing symmetry is then the condition that the scattering $S$-matrix for a process with an in-state that includes particle $A$, and some out-state, has \textit{the same} analytic form as the ``crossed'' versions of the original process, with an outgoing anti-particle $\bar{A}$. While in the original process, the particle appearing in the in-state carries positive energy, in the crossed process, the anti-particle $\bar{A}$ appearing in the out-state carries negative energy.
However, crossing symmetry allows one to use the same analytic $S$-matrix element to also calculate the process with an outgoing anti-particle $\bar{A}$ in its physical kinematic regime. In the presence of crossing symmetry, a single analytic function provides the $S$-matrix for several different processes in different regions of complexified momentum space.  For massless particles, under crossing, 
\bea
 {\rm particle} \; & \leftrightarrow & \; {\rm antiparticle} \nonumber \\
 {\rm incoming} \; & \leftrightarrow & \; {\rm outgoing} \nonumber \\
 {\rm helicity} \; h \; & \leftrightarrow & \; - h \nonumber \\
 p^\mu \; & \leftrightarrow & \; - p^\mu \nonumber
\eea
Since the $S$-matrix for electric-magnetic scattering processes does not obey crossing symmetry, one can not describe different processes using the same $S$-matrix element. Nevertheless, we can still use a {\it crossing transformation} to translate the problem formulated in \emph{in-out} language into the \emph{out-out} language, which is the conventional choice of the spinor-helicity community. This is possible because, as can be seen from Eq.~(\ref{eq:redef}), the LG transformation of an $S$-matrix involving incoming states with helicities $h_i$ and pairwise helicities $q_{ij}$ is the \textit{same} as that of an $S$-matrix with outgoing states with helicities $-h_i$ and pairwise helicities $q_{ij}$.

Consequently, we are free to construct $S$-matrices in the \emph{out-out} formalism, as long as we keep working in the same kinematic regime of the original \emph{in-out} $S$-matrix. 
Furthermore, even in the \emph{out-out} formalism, we consider pairwise helicities $q_{ij}$ only for pairs of states which are both in the initial state or both in the final state for a given physical process. 

\subsection{Constructing the electric-magnetic $S$-matrix: spinor-helicity cheat sheet}

We are now ready to formulate general rules for constructing electric-magnetic $S$-matrix elements. As usual in the amplitudes program, the spinor-helicity variables are the basic building blocks. The main novelty is the appearance of the pairwise spinor-helicity variables, needed to capture the additional pairwise LG phase in the $S$-matrix,  in addition to the ordinary ones. As usual, we will assign helicity weights (or for massive particles $SU(2)$ quantum numbers) to each spinor-helicity variable, as well as a {\it separate} pairwise helicity weights to each pairwise spinor-helicity variable. We will require that the helicity weights under each individual particle as well as the pairwise helicity weights are matched for both the initial and the final states. Of course only the diagonal Lorentz transformation (where each particle and each pair of particles are  transformed simultaneously) is physical. However, as is common in the amplitudes approach, as a book-keeping tool we can pretend that helicity and pairwise helicity transformations can be performed independently on each particle/pair of particles, which will make the construction of the properly transforming $S$-matrix particularly easy. Hence for the pairwise helicity variable we assign {\it only} the pairwise helicity (and no ordinary helicities), even though these pairwise spinor-helicity variables are constructed as a function of the ordinary helicity variables, and in some limits they even coincide with one of the ordinary spinor-helicity variables.\footnote{In the massless limit, the regular LG phase coincides with the pairwise phase, and LG weights of some of the regular variables are used to match the regular LG weights, while the rest are used to saturate the pairwise LG weight. } 

These rules are summarized by the following equations.
\bea
&& S \left( \omega^{-1} \vert i \rangle, \omega \vert i ] \right) = \omega^{2 h_i} S \left( \vert i \rangle, \vert i ] \right), \;\; {\rm for} \; \forall i \label{eq:EM_S_matrix_regular_LG} \\
&& S \left( \omega^{-1} \vert p^{\flat +}_{ij} \rangle,  \omega \vert p^{\flat +}_{ij} ],  \omega \vert p^{\flat -}_{ij} \rangle,  \omega^{-1} \vert p^{\flat -}_{ij} ] \right) = \omega^{-2 q_{ij}} S \left(  \vert p^{\flat +}_{ij} \rangle , \vert p^{\flat +}_{ij} ],  \vert p^{\flat -}_{ij} \rangle,  \vert p^{\flat -}_{ij} ] \right) \;\; {\rm for} \; \forall \; {\rm pair} \; \{i,j\}, \nonumber \\ \label{eq:EM_S_matrix_pairwise_LG}
\eea
where $\omega$ represents the LG weight +1/2.
The resulting rules for the full set of charge assignments of the spinor-helicity variables are presented in Table~\ref{tab:cheat}, which summarizes the different LG weights of the regular and pairwise spinor-helicity variables, as well as the overall weights of the amplitude implied from Eq.~(\ref{eq:EM_S_matrix_regular_LG}) and (\ref{eq:EM_S_matrix_pairwise_LG}). 

\begin{table}[htb]
\centering
\begin{tabular}{@{}cccc@{}}
\textbf{}                                                         & $U(1)_i$   &$SU(2)_i$ & $U(1)_{ij}$ \\ \midrule
\addlinespace[1em]
Required weight~~~~~& $ h_i$                     & $\mathbf{S}_i$            & $\text{-}q_{ij}$                   \\
\addlinespace[1em]
                                           $\ket{i}_\alpha,\,\sbra{i}_{\dot{\alpha}}$  ~~~~                      & $\text{-}\frac{1}{2}\,,\,\frac{1}{2}$ & $-$                         & $-$                           \\
\addlinespace[1em]
                        $\bra{\mathbf{i}}^{I;\alpha}$    ~~~~                                       & $-$                        & $\square$                 & $-$                         \\
\addlinespace[1em]
$\ket{p^{\flat+}_{ij}}_\alpha,\,\sbra{p^{\flat+}_{ij}}_{\dot{\alpha}}$                ~~~~                                                   & $-$                        & $-$                         & $\text{-}\frac{1}{2}\,,\,~\frac{1}{2}~$  \\
\addlinespace[1em]
$\ket{p^{\flat-}_{ij}}_\alpha,\,\sbra{p^{\flat-}_{ij}}_{\dot{\alpha}}$       ~~~~                                                             & $-$                        & $-$                          & $\frac{1}{2}\,,\,\text{-}\frac{1}{2}$ 
\end{tabular}
\caption{LG weights of the standard and pairwise spinor-helicity variables, as well as the overall weight required by Eq.~(\ref{eq:EM_S_matrix_regular_LG}) and (\ref{eq:EM_S_matrix_pairwise_LG}).}
\label{tab:cheat}
\end{table}


\subsection{First examples}

To illustrate the construction of electric-magnetic $S$-matrix elements, let us work out a few examples.\\ \quad\\
\noindent \textbf{(1) Massive fermion decaying to massive fermion  + massless scalar, $q=-1$.}\\
In this case we need to use one massive spinor for the decaying fermion and one massive spinor for the final fermion.  This gives us two spinor indices that should be contracted with pairwise spinors. 
Note that in general, the number of pairwise spinors is not completely fixed by the LG: only the difference $n^{-}_{23}-n^{+}_{23}$ between the number of pairwise spinors with weight $\frac{1}{2}$ and $-\frac{1}{2}$ is fixed to be $-2q_{23}$. In our case we need a total of 2 spinor indices and so $n^{+}_{23}=2,\,n^{-}_{23}=0$. The $S$-matrix is then
\begin{eqnarray} 
S\left(\mathbf{1}^{s=1/2}\,|\,\mathbf{2}^{s=1/2},3^{0}\right)_{q_{23}=-1}~&\sim&~\bk{p^{\flat-}_{23}\,\mathbf{1}}\bk{p^{\flat-}_{23}\,\mathbf{2}}\,,
\label{eq:first_eg}
\end{eqnarray}
up to a LG invariant.\footnote{In principle, there are other ``legally'' acceptable expressions such as $\left[ p^{\flat+}_{23} \mathbf{1} \right] \left[ p^{\flat+}_{23} \mathbf{2} \right]$ or $\left[ p^{\flat+}_{23} \mathbf{1} \right] \left\langle p^{\flat-}_{23} \mathbf{2} \right\rangle$ or $\left\langle p^{\flat-}_{23} \mathbf{1} \right\rangle \left[ p^{\flat+}_{23} \mathbf{2} \right]$. However, using the Dirac equations for the massive variable, $p_{\alpha \dot{\alpha}} \tilde{\lambda}^{\dot{\alpha} I} = m \lambda^I_\alpha$ and $p_{\alpha \dot{\alpha}} \lambda_{\alpha I} = - m \tilde{\lambda}^I_{\dot{\alpha}}$, one can check that these are equivalent to Eq.~(\ref{eq:first_eg}) up to LG invariants.}
\\ \quad\\
\noindent \textbf{(2) Massive scalar decaying to massive scalar + massless vector, $q=-1$.}\\
In this case we need to use two regular spinor-helicity variables for the helicity of the vector, as well as two pairwise spinors for the $q_{23}=-1$ of the final state. The $S$-matrix elements for helicity $\pm1$ vectors are then
\begin{eqnarray} 
S\left(\mathbf{1}^{s=0}\,|\,\mathbf{2}^{s=0},3^{+1}\right)_{q_{23}=-1}~&\sim&~\,\bks{p^{\flat+}_{23}\,3}^2\sim\bmk{p^{\flat-}_{23}}{2}{3}^2\, ,
\end{eqnarray}
up to a LG invariant. On the other hand, there is no way to write a LG covariant expression for $S\left(\mathbf{1}^{s=0}\,|\,\mathbf{2}^{s=0},3^{-1}\right)_{q_{23}=-1}$. We will see later that this is a particular example of a more general LG \textit{selection rule}.\\ \quad\\
\noindent \textbf{(3) Massive vector decaying to two different massless fermions, $q=-2$.}\\
In this case we need to use 2 massive spinors for the vector and one regular spinor-helicity variable for each fermion, as well as four pairwise spinors for the $q_{23}=-2$ of the out state. The $S$-matrix for opposite helicity fermions is then 
\begin{eqnarray} 
S\left(\mathbf{1}^{s=1}\,|\,2^{-1/2},3^{+1/2}\right)_{q_{23}=-2}~&\sim&~\bk{2 p^{\flat-}_{23}}\bks{p^{\flat+}_{23}\,3}\,\bk{\mathbf{1}\,p^{\flat-}_{23}}^2\, .
\end{eqnarray}
up to a LG invariant. Note that the $S$-matrix for same helicity fermions\footnote{In the all-\textit{outgoing} sense.} is forbidden in this case, due to the fact that $\bk{p^{\flat-}_{23}3}=\bks{p^{\flat+}_{23}2}=0$. This is our second encounter with a LG selection rule.\\ \quad\\
\noindent \textbf{(4) Massive vector decaying to two different massless fermions, $q=-1$.}\\
In this case we need to use 2 massive spinors for the vector and one regular spinor-helicity variable for each fermion, as well as four pairwise spinors for the $q_{23}=-1$ of the out state. Note that unlike the previous examples, here the total number of pairwise spinors is \textit{not} given by $-2q_{23}$. This is because there are four spinor indices from the standard spinors that need to be contracted, so that $n^{+}_{23}+n^{-}_{23}=4$. Pairwise LG, on the other hand, implies $n^{+}_{23}-n^{-}_{23}=-2q_{23}=2$, and so we have $n^{+}_{23}=3,\,n^{-}_{23}=1$. The $S$-matrix for positive helicity fermions is then
\begin{eqnarray} 
S\left(\mathbf{1}^{s=1}\,|\,2^{-1/2},3^{-1/2}\right)_{q_{23}=-1}~&\sim&~\bk{2 p^{\flat-}_{23}}\bk{p^{\flat+}_{23}\,3}\,\bk{\mathbf{1}\,p^{\flat-}_{23}}^2\, .
\end{eqnarray}
up to a LG invariant. Note that the $S$-matrix for $h_2=-h_3=1/2$  is forbidden in this case, due to the fact that $\bks{p^{\flat-}_{23}3}=0$.

\subsection{All electric-magnetic 3-point $S$-matrix elements}\label{sec:3pt}
The examples above give us a flavor of how to construct electric-magnetic $S$-matrix elements up to LG invariants. In the case of 3-point $S$-matrix elements, we can make the discussion even more concrete and write down systematic expressions and selection rules for electric-magnetic $S$-matrix elements. These are modifications of the general 3-point amplitudes derived in \cite{Arkani-Hamed:2017jhn}, when the three scattering particles can have magnetic charge. Without loss of generality, we choose one massive particle (that may be a dyon) in the incoming state, and two particles (that may also be dyons) in the outgoing state. Note that our expressions extend the ones presented in \cite{Arkani-Hamed:2017jhn} to the case of electric-magnetic scattering, and reduce to them when $q=0$ for the outgoing states. Below, whenever we call a particle ``dyon'', we mean that it may, or may not, have a magnetic charge. In all our cases, the decaying particle may be any kind of ``dyon''.

$\bullet$ \textit{Incoming massive particle, two outgoing massive particles}

In this case the $S$-matrix is the contraction of the massive part (in the notation of \cite{Arkani-Hamed:2017jhn})
\begin{eqnarray}
\left({\bra{\mathbf{1}}^{2s_1}}\right)^{\left\{\alpha_1\ldots\alpha_{2s_1}\right\}}\left({\bra{\mathbf{2}}^{2s_2}}\right)^{\left\{\beta_1\ldots\beta_{2s_2}\right\}}{\left(\bra{\mathbf{3}}^{2s_3}\right)}^{\left\{\gamma_1\ldots\gamma_{2s_3}\right\}}
\end{eqnarray}
with a massless part involving the pairwise spinors $\ket{w}_{\alpha}\equiv\ket{p^{\flat-}_{23}}_{\alpha}$ and $\ket{r}_{\alpha}\equiv\ket{p^{\flat+}_{23}}_{\alpha}$ (with pairwise helicities $\pm\frac{1}{2}$), which saturates the pairwise LG transformation. The most general expression is
 \begin{eqnarray}\label{eq:3mas}
&&S^{q}_{\left\{\alpha_1,\ldots,\alpha_{2s_1}\right\}\left\{\beta_1,\ldots,\beta_{2s_2}\right\}\left\{\gamma_1,\ldots,\gamma_{2s_3}\right\}}~=~\sum_{i=1}^C~a_i\,{\left(\ket{w}^{\hat{s}-q}\,\ket{r}^{\hat{s}+q}\right)}_{\left\{\alpha_1,\ldots,\alpha_{2s_1}\right\}\left\{\beta_1,\ldots,\beta_{2s_2}\right\}\left\{\gamma_1,\ldots,\gamma_{2s_3}\right\}}\, ,\nonumber\\
\end{eqnarray}
 where $\hat{s}=s_1+s_2+s_3$, $C$ counts all the possible ways to group the spinors into $\alpha,\,\beta$ and $\gamma$ indices, and $q=q_{23}=e_2g_3-e_3g_2$. Since both exponents have to be non-negative integers, we get a \textit{selection rule}: 
  \begin{eqnarray}\label{eq:sel1}
|q|\leq\hat{s}\ .
\end{eqnarray}
We can also check that Eq.~(\ref{eq:3mas}) reduces to the standard expression from \cite{Arkani-Hamed:2017jhn} for $q=0$. To see this, note that 
  \begin{eqnarray}
{\left(\,\ket{w}\,\ket{r}\,\right)}_{\left\{\alpha\beta\right\}}~&\sim&~\mathcal{O}_{\left\{\alpha\beta\right\}}~\equiv~\left(p_2\right)_{\left\{\alpha\dot{\gamma}\right.} \left(p_3\right)^{~~\,\,\dot{\gamma}}_{\left.\beta\right\}}\nonumber\\
{\left(\,\ket{w}\,\ket{r}\,\right)}_{\left[\alpha\beta\right]}~&\sim&~\varepsilon_{\alpha\beta}\, .
\end{eqnarray}
where the two index tensors $\mathcal{O}_{\left\{\alpha\beta\right\}}$ were defined in  \cite{Arkani-Hamed:2017jhn}. This can be seen from Eq.~(\ref{eq:ppmeqinv}), i.e. 
  \begin{eqnarray}
\left(p_2\right)_{\left\{\alpha\dot{\gamma}\right.} \left(p_3\right)^{~~\,\,\dot{\gamma}}_{\left.\beta\right\}}~=~\frac{E^c_2 + E^c_3}{2p_c}\,\left(p^{\flat+}_{23}\right)_{\left\{\alpha\dot{\gamma}\right.} \left(p^{\flat-}_{23}\right)^{~~\,\,\dot{\gamma}}_{\left.\beta\right\}}~=~(E^c_2 + E^c_3)\,{\left(\,\ket{w}\,\ket{r}\,\right)}_{\left\{\alpha\beta\right\}}\,.\nonumber\\
\end{eqnarray}
When $q=0$, we get Eq.~(4.27) of \cite{Arkani-Hamed:2017jhn},
 \begin{eqnarray}\label{eq:3mas2}
&&S^{0}_{\left\{\alpha_1,\ldots,\alpha_{2s_1}\right\}\left\{\beta_1,\ldots,\beta_{2s_2}\right\}\left\{\gamma_1,\ldots,\gamma_{2s_3}\right\}}~=~\sum_{i=0}^1~\tilde{a}_i\,{\left(\mathcal{O}^{\hat{s}-i}\varepsilon^i\right)}_{\left\{\alpha_1,\ldots,\alpha_{2s_1}\right\}\left\{\beta_1,\ldots,\beta_{2s_2}\right\}\left\{\gamma_1,\ldots,\gamma_{2s_3}\right\}}\, .\nonumber\\
\end{eqnarray}

$\bullet$ \textit{Incoming massive particle, outgoing massive particle + massless particle; unequal mass case.}\\

This is the electric-magnetic version of the two massive, one massless $S$-matrix from \cite{Arkani-Hamed:2017jhn}. 
In this case the $S$-matrix is the contraction of the massive part 
\begin{eqnarray}
{\left(\bra{\mathbf{1}}^{2s_1}\right)}^{\left\{\alpha_1\ldots\alpha_{2s_1}\right\}}{\left(\bra{\mathbf{2}}^{2s_2}\right)}^{\left\{\beta_1\ldots\beta_{2s_2}\right\}}\, ,
\end{eqnarray}
with the massless part constructed from two ``regular'' spinors:
\begin{eqnarray}
\left(\ket{u}_\alpha,\ket{v}_\alpha\right)~=~\left(\ket{3}_\alpha,{|\,2\sket{3}}_\alpha\right)\, ,
\end{eqnarray}
with regular LG weights $\mp \frac{1}{2}$, as well as the pairwise spinors
\begin{eqnarray}
\left(\ket{w}_\alpha,\ket{r}_\alpha\right)~=~\left(\ket{p^{\flat-}_{23}}_\alpha,\ket{p^{\flat+}_{23}}_\alpha\right)\, ,
\end{eqnarray}
with pairwise LG weights $\pm\frac{1}{2}$. Note that $|\,2\,|p^{\flat-}_{23}]_\alpha$  is nothing but a LG invariant times $\ket{p^{\flat+}_{23}}_\alpha$.

The general massive 3-point $S$-matrix for an initial spin $s_1$ particle and an final spin $s_2$ particle is then
\begin{eqnarray}\label{eq:3uneq}
S^{h,q,\text{ unequal}}_{\left\{\alpha_1,\ldots,\alpha_{2s_1}\right\}\left\{\beta_1,\ldots,\beta_{2s_2}\right\}}~&=&~\sum_{i=1}^C~\sum_{j,k}\,a_{ijk}\,\bk{ur}^{\text{max}(j+k,0)}\,\bk{vw}^{\text{max}(-j-k,0)}\,\nonumber\\&&{\left(\ket{u}^{\frac{\hat{s}}{2}-h-j}\ket{v}^{\frac{\hat{s}}{2}+h+k}\ket{w}^{\frac{\hat{s}}{2}-q+j}\ket{r}^{\frac{\hat{s}}{2}+q-k}\right)}_{\left\{\alpha_1,\ldots,\alpha_{2s_1}\right\}\left\{\beta_1,\ldots,\beta_{2s_2}\right\}}\, ,\nonumber\\
\end{eqnarray}
where $\hat{s}=s_1+s_2$, and $q=q_{23}=e_2g_3-e_3g_2$. Again $C$ is the number of distinct tensor structures. 
The $j$ and $k$ sums are over values that give non-negative exponents. In particular, they are in the intervals $-\frac{\hat{s}}{2}+q\leq j \leq \frac{\hat{s}}{2}-h$ and $-\frac{\hat{s}}{2}-h\leq k \leq \frac{\hat{s}}{2}+q$ . These intervals exist only if $|h+q|\leq\hat{s}$, which gives us a \textit{selection rule}. In particular, 
\begin{eqnarray}
s_1~=~s_2~=~0~~\rightarrow~~h~=~-q\,.
\end{eqnarray}
$\bullet$ \textit{Incoming massive particle, outgoing massive particle + massless particle; equal mass case.}\\
\vspace*{0.01cm}

When the two masses are equal, we know that $\langle u v \rangle \propto p_2 \cdot p_3 = 0$, hence, $u$ and $v$ are parallel.
For constructing the $S$-matrix, therefore, we use only one of the two, say $\ket{u}$. However, the ratio $x$ of the two is defined via\footnote{An alternative expression for this $x$-factor can be written as \cite{Arkani-Hamed:2017jhn}
\begin{eqnarray}\label{eq:exf}
x~=~\frac{\bmk{\zeta}{2}{3}}{m\bk{\zeta3}}\, ,
\end{eqnarray}
where $\bra{\zeta}$ is an arbitrary spinor which drops out of any physical calculation.
}
\begin{eqnarray}
m\,x \ket{u}~=~\ket{v}\, ,
\end{eqnarray}
and carries regular helicity of $+1$ for the particle 3, and can be used to satisfy the regular helicity weight of the $S$-matrix. Similarly, $\langle w r \rangle = 0$ and we have the relation
\bea
\bk{ur}^2\,x \ket{w}~\sim~\,\ket{r}\, ,
\eea
up to an overall LG invariant. Overall, the $S$-matrix is then constructed using $x, \ket{u}_\alpha, \ket{w}_\alpha$ and $\epsilon_{\alpha \beta}$. 
A solution consistent with the regular/pairwise helicity weight and the number of required spinor indices is found to be
\begin{eqnarray}
S^{h,q, {\rm equl}}_{\{ \alpha_1 \ldots \alpha_{2s_1}  \} \{ \beta_1 \ldots \beta_{2s_2} \} }~&=&~\sum_{i=1}^C \sum_{j}\,\sum_{k=-j}^j~x^{h+q+j}\,\bk{ur}^{\text{max}\left[2q+j-k,0\right]}\,\bk{vw}^{\text{max}\left[-2q-j+k,0\right]}\,\cdot\, \nonumber\\
&&~~~~~~~~~~~\left( \ket{u}^{j+k} \ket{w}^{j-k} \epsilon^{\hat{s} - j} \right)_{\left\{ \alpha_1 \ldots \alpha_{2s_1} \right\} \left\{ \beta_1 \ldots \beta_{2s_2}\right\} } ,\nonumber \\
\end{eqnarray}
where the $j$ sum extends over $0\leq j \leq \hat{s}$. Note that while the powers of $u, w, \epsilon$ have to be non-negative integers, there is no such requirement for the power of $x$.

$\bullet$ \textit{Incoming massive particle, two outgoing massless particles}\\

In this case the $S$-matrix is the contraction of the massive part 
\begin{eqnarray}
{\left(\bra{\mathbf{1}}^{2s}\right)}^{\left\{\alpha_1\ldots\alpha_{2s}\right\}}
\end{eqnarray}
with a massless part involving the regular spinors $\ket{u}_\alpha=\ket{2}_\alpha,\,\ket{v}_\alpha=\ket{3}_\alpha$ and the pairwise spinors $\ket{w}_\alpha=\ket{p^{\flat-}_{23}}_\alpha$ and $\ket{r}_\alpha=\ket{p^{\flat+}_{23}}_\alpha$. The most general expression is
 
 \begin{eqnarray}
&&S^{q}_{\left\{\alpha_1,\ldots,\alpha_{2s}\right\}}~=~\sum_{ij}~a_{ij}~{\left(\ket{u}^{s/2-i-\Delta}~\ket{v}^{s/2-j+\Delta}~\ket{w}^{s/2+j-q}~\ket{r}^{s/2+i+q}\right)}_{\left\{\alpha_1,\ldots,\alpha_{2s}\right\}}\,\cdot\nonumber\\
&&\bks{uv}^{\text{max}\left[\Sigma+(s-i-j)/2\,,\,0\right]}\,\bk{uv}^{\text{max}\left[-\Sigma-(s+i+j)/2\,,\,0\right]}~{\left(\bk{uw}\bks{vr}\right)}^{\frac{1}{2}\text{max}\left[i-j\,,\,0\right]}~ {\left(\bks{uw}\bk{vr}\right)}^{\frac{1}{2}\text{max}\left[j-i\,,\,0\right]}\, ,\nonumber\\
\end{eqnarray}
with $\Sigma=h_2+h_3,\,\Delta=h_2-h_3$. Again $q=q_{23}=e_2g_3-e_3g_2$, and the $i$ and $j$ sums are over values in the intervals ${-s/2-q\leq\, i\,\leq s/2-\Delta}$ and ${-s/2+q\leq\, j\,\leq s/2+\Delta}$, such that all of the exponents are non-negative integers. These intervals exists only when $|\Delta-q|\leq s$, which gives us another \textit{selection rule}. In the non-magnetic $q=0$ case, this gives us the same selection rule as \cite{Arkani-Hamed:2017jhn}. In particular, for a spin $s$ coupling to $h_2=-h_3$, we have
\begin{eqnarray}
&&\text{For }q=0\,:\nonumber\\[10pt]
&&s~=~0~~\rightarrow~~h_2~=~h_3~=~0\nonumber\\[10pt]
&&s~=~1~~\rightarrow~~|h_2\,-\,h_3|~\leq~1~~\rightarrow~~|h_2|=|h_3|\leq1/2\nonumber\\[10pt]
&&s~=~2~~\rightarrow~~|h_2\,-\,h_3|~\leq~2~~\rightarrow~~|h_2|=|h_3|\leq1~~~~~~~~~~~~~~~~~~~~~~~~~~~~~~~~~~~~~~~~~~~\,,
\end{eqnarray}
in other words, massless particles with $|h|>\frac{1}{2}$ cannot couple to a Lorentz covariant conserved current, and massless particles with $|h|>1$ cannot couple to a conserved stress tensor. 
For $q\neq 0$, the situation is even more restrictive. For example, when $|q|=\frac{1}{2}$ we have
\begin{eqnarray}
&&\text{For }q=\pm1/2\,:\nonumber\\[10pt]
&&s~=~0~~\rightarrow~~\text{forbidden}\nonumber\\[10pt]
&&s~=~1~~\rightarrow~~|h_2\,-\,h_3\mp1/2|~\leq~1~~\rightarrow~~|h_2|~=~|h_3|~=~0~~~~\text{or}~~h_2=-h_3=\pm1/2\nonumber\\[10pt]
&&s~=~2~~\rightarrow~~|h_2\,-\,h_3\mp1/2|~\leq~2~~\rightarrow~~|h_2|~=~|h_3|\,\leq\,1/2~~\text{or}~~h_2=-h_3=\pm1\,.\nonumber\\
\end{eqnarray}
We see that for $|q|=1/2$ the selection rule is more restrictive than in the $q=0$ case, since it discards the $h_2=-h_3=-qs$ option.

\section{Partial Wave Decomposition for $2\rightarrow 2$ Electric-Magnetic $S$-matrix}\label{sec:twotwo}

Following \cite{Arkani-Hamed:2017jhn} and \cite{Jiang:2020sdh},  we can now perform a relativistic partial wave decomposition for $2\rightarrow 2$ electric-magnetic $S$-matrix elements\footnote{For a complementary approach to mapping all possible spinor structures for 4-point non-magnetic amplitudes, see \cite{Durieux:2020gip}}. In a Poincar\'e invariant setting, the partial wave decomposition is nothing but the expansion in a complete eigenbasis of the Casimir operator $W^2$, where $W^\mu$ is the Pauli-Lubanski operator defined by
 \begin{eqnarray}
W^\mu~\equiv~\frac{1}{2}\,\epsilon_{\mu\nu\rho\sigma}\,P^\nu \,M^{\rho\sigma}\,.
\end{eqnarray}
In the above expression $P^\nu$ is the momentum operator and $M^{\rho\sigma}$ is the Lorentz generator. The eigenvalues of $W^2$ are given by $-P^2\,J\,(J+1)$ where $J$ is the total angular momentum, so clearly this is the relativistic version of a partial wave decomposition. The operators $P^\mu,\,M^{\mu\nu}$ and $W_\mu$ act on the \textit{amplitude} or parts of it. In particular, we will make use of their representation as differential operators acting in spinor-helicity space \cite{Witten:2003nn}. In the non-magnetic case and for massless particles, these are given by \cite{Witten:2003nn,Conde:2016izb}
 \begin{eqnarray}\label{eq:lorop}
\left(\sigma_\mu\right)_{\alpha\dot{\alpha}}\,P^\mu~&\equiv&~P_{\alpha\dot{\alpha}}~=~\sum_i\,\ket{i}_\alpha \sbra{i}_{\dot{\alpha}}\nonumber\\
\left(\sigma_{\mu\nu}\right)_{\alpha\beta}\,M^{\mu\nu}~&\equiv&~M_{\alpha\beta}~=~i\,\sum_i\,\ket{i}_{\left\{\alpha\right.} \,\frac{\partial}{\partial\bra{i}^{\left.\beta\right\}}}\nonumber\\
\left(\bar{\sigma}_{\mu\nu}\right)_{\dot{\alpha}\dot{\beta}}\,M^{\mu\nu}~&\equiv&~\tilde{M}_{\dot{\alpha}\dot{\beta}}~=~i\,\sum_i\,\sbra{i}_{{\left\{\dot{\alpha}\right.}} \,\frac{\partial}{\partial\sket{i}^{{\left.\dot{\beta}\right\}}}}\, ,
\end{eqnarray}
where the sum $i$ is over a collection of particles. In the $2\rightarrow 2$ case we are interested in the total angular momentum of particles 1 and 2, and so the sum will be over $i=1,2$. The generalization of Eq.~(\ref{eq:lorop}) for massive particles is straightforward \cite{Conde:2016izb,Guevara:2018wpp}: we bold the spinors and contract their $SU(2)$ LG indices. The Casimir operator $W^2$ is then expressible as \cite{Conde:2016izb,Jiang:2020sdh}
 \begin{eqnarray}\label{eq:cas}
W^2~=~\frac{P^2}{8}\,\left[\text{Tr}\left(M^2\right)\,+\,\text{Tr}\left(\tilde{M}^2\right)\right]\,-\,\frac{1}{4}\,\text{Tr}\left(M\,P\,\tilde{M}\,P^{\text{T}}\right)\, .
\end{eqnarray}
Eq.~(\ref{eq:lorop}) can be straightforwardly generalized to our electric-magnetic case by treating the regular and pairwise spinors on the same footing:
 \begin{eqnarray}\label{eq:loropgen}
\left(\sigma_{\mu\nu}\right)_{\alpha\beta}\,M^{\mu\nu}~&\equiv&~M_{\alpha\beta}~=~i\,\left[\sum_i\,\ket{i}_{\left\{\alpha\right.} \,\frac{\partial}{\partial\bra{i}^{\left.\beta\right\}}}\,+\,\sum_{i>j,\pm}\,\ket{p^{\flat\pm}_{ij}}_{\left\{\alpha\right.} \,\frac{\partial}{\partial\bra{p^{\flat\pm}_{ij}}^{\left.\beta\right\}}}\right]\nonumber\\
\left(\bar{\sigma}_{\mu\nu}\right)_{\dot{\alpha}\dot{\beta}}\,M^{\mu\nu}~&\equiv&~\tilde{M}_{\dot{\alpha}\dot{\beta}}~=~i\,\left[\sum_i\,\sbra{i}_{{\left\{\dot{\alpha}\right.}} \,\frac{\partial}{\partial\sket{i}^{{\left.\dot{\beta}\right\}}}}\,+\,\,\sum_{i>j,\pm}\,\sbra{p^{\flat\pm}_{ij}}_{\left\{\dot{\alpha}\right.} \,\frac{\partial}{\partial\sket{p^{\flat\pm}_{ij}}^{\left.\dot{\beta}\right\}}}\right]\, ,
\end{eqnarray}
where the sum is now over all pairs as well as individual particles in the state. It is easy to see that
 \begin{eqnarray}\label{eq:Wsq}
W^2\,\bk{12}~=~W^2\,\bk{p^{\flat\pm}_{12}2}~=~W^2\,\bk{p^{\flat\pm}_{12}1}~=~W^2\,\bk{p^{\flat\pm}_{12}p^{\flat\mp}_{12}}~=~0\, ,
\end{eqnarray}
with $W^2$ the Casimir associated with particles 1 and 2 and defined via Eq.~(\ref{eq:loropgen}). Similarly,
 \begin{eqnarray}\label{eq:Wsq2}
W^2\,\ket{\,1^{\color{white}\flat}}_{\left\{\alpha\right.}\ket{p^{\flat-}_{12}}_{\left.\beta\right\}}~=~
-s\,1(1+1)\,\ket{\,1^{\color{white}\flat}}_{\left\{\alpha\right.}\ket{p^{\flat-}_{12}}_{\left.\beta\right\}}\, .
\end{eqnarray}
In other words, the eigenfunctions of $W^2$ are combinations of regular and pairwise spinors with symmetrized spinor indices. The eigenvalues are $-s\,j\,(j+1)$ where $j$ is just the number of uncontracted spinor indices, divided by 2.  This is the same conclusion as in ref.~\cite{Jiang:2020sdh}, only with the inclusion of of pairwise spinors in the definition of $W^2$. It is now natural to expand the $S$-matrix in a complete eigenbasis of $W^2$ with eigenfunctions
 \begin{eqnarray}
W^2\,\,\mathcal{B}^{J}~=~-s\,J\,(J+1)\,\,\mathcal{B}^{J}\, .
\end{eqnarray}
Following \cite{Jiang:2020sdh}, we call the $\mathcal{B}^J$ \textit{basis amplitudes}. 
The most general expansion then reads
 \begin{eqnarray}
 S_{12\rightarrow34}~=~\mathcal{N}\,\sum_{J}\,(2J+1)\,\mathcal{M}^J(p_c)\,\mathcal{B}^{J}\, ,
\end{eqnarray}
where $\mathcal{N}\equiv\sqrt{8\pi s}\,$ is a normalization factor and $\mathcal{M}^J(p_c)$ are coefficients\footnote{We also added the factor $(2J+1)$ as part of normalization so that the partial wave unitarity equation is expressed in a simple form in terms of $\mathcal{M}^J(p_c)$ Eq.~(\ref{eq:unitrel}).} satisfying
 \begin{eqnarray}
W^2_{12}\,\,\mathcal{M}^{J}(p_c)~&=&~W^2_{34}\,\,\mathcal{M}^{J}(p_c)~=~0\, .
\end{eqnarray}
The eigenfunctions $\mathcal{B}^{J}$ are then nothing but symmetrized products of spinors,
 \begin{eqnarray} \label{eq:Def_of_S_matrix_CG_M}
\mathcal{B}^{J}~=~~C^{J;\,\text{in}}_{\left\{\alpha_1,\ldots,\alpha_{2j}\right\}}\,C^{J;\,\text{out};~\left\{\alpha_1,\ldots,\alpha_{2j}\right\}}\, ,
\end{eqnarray}
where
 \begin{eqnarray}
W^2_{12}\,\,C^{J;\,\text{in}}_{\left\{\alpha_1,\ldots,\alpha_{2J}\right\}}~&=&~-s\,J\,(J+1)\,C^{J;\,\text{in}}_{\left\{\alpha_1,\ldots,\alpha_{2J}\right\}}\, \nonumber\\[10pt]
W^2_{34}\,\,C^{J;\,\text{out};~\left\{\alpha_1,\ldots,\alpha_{2J}\right\}}~&=&~-s\,J\,(J+1)\,C^{J;\,\text{out};~\left\{\alpha_1,\ldots,\alpha_{2J}\right\}}\,.
\end{eqnarray}
In the above expression $W^2_{12}$ and $W^2_{34}$ are the Casimir operators associated with particles 1,2 and 3,4, respectively. The coefficient functions $\mathcal{M}^{J}(p_c)$ are angular momentum singlets, and so they can only depend on the energy scale of the scattering, given by the COM momentum $p_c\,$. Inspired by the Wigner-Eckart theorem, we call them ``reduced matrix elements''. They contain the dynamical information of the scattering process, as opposed to the angular dependence that is fixed for every partial wave. The coefficients $C^{J;\,\text{in/out}}$, on the other hand, are generalized Clebsch-Gordan coefficients \cite{Jiang:2020sdh}.\footnote{To be more precise, our $C^{J;\,\text{in/out}}$ are not really {\it coefficients}, they are $SL(2, \mathbb{C})$ tensors. The generalized Clebsch-Gordan coefficients defined in \cite{Jiang:2020sdh} is given in terms of our $C^{J;\,\text{in/out}}$ by $C^{J;\,\text{in/out}; \{ \alpha_1 \ldots, \alpha_{2J} \}} \lambda_{\alpha_1}^{I_1} \cdots \lambda_{\alpha_{2J}}^{I_{2J}}$.} These coefficients are completely fixed by group theory, and we can easily find them using an elegant trick from \cite{Arkani-Hamed:2017jhn,Jiang:2020sdh}. Simply put, the Clebsch-Gordan coefficient connecting the particles $i$ and $j$ to the total angular momentum $J$ is directly extracted from the 3-point  $S$-matrix element with the particles $i$ and $j$ and a massive, spin $J$ particle. 
For example, if 1 and 2 are two massive scalar dyons with $q_{12}=-1$, the corresponding 3-point $S$-matrix element is
   \begin{eqnarray}\label{eq:3ptex1}
S\left(1^0,2^0\,|\,\mathbf{3}^J\right)_{q_{12}=-1}~=~a\,\bk{\mathbf{3}\,p^{\flat-}_{12}}^{J+1}\,\bk{\mathbf{3}\,p^{\flat+}_{12}}^{J-1}\, .
\end{eqnarray}
Since there is only one relevant tensor structure for this $S$-matrix (see Eq.~(\ref{eq:3mas})), we have only one coefficient $a$. This will change when we include non-scalar particles --- for example with a massive fermion $f$ and a scalar there are two possible tensor structures, depending on which spinor is contracted with $\sket{\mathbf{f}}$. The corresponding generalized Clebsch-Gordan part can be directly read off from this 3-point $S$-matrix element by stripping off the spinors $\bra{\mathbf{3}}^{\alpha}$ corresponding to the massive spin $J$, 
   \begin{eqnarray}\label{eq:3ptex2}
{\left(C^{J; \,\text{in}}_{0,0,-1}\right)}_{\left\{\alpha_1,\ldots,\alpha_{2J}\right\}}~=~{\left(\ket{p^{\flat-}_{12}}^{J+1}\,\ket{p^{\flat+}_{12}}^{J-1}\right)}_{\left\{\alpha_1,\ldots,\alpha_{2J}\right\}}\, ,
\end{eqnarray}
where the subscript $(0,0,-1)$ indicates $(s_1,s_2,q_{12})$ and we have normalized away the $a$ coefficient.

\section{Fermion-Monopole Scattering: Lowest Partial Wave and Helicity Flip}\label{sec:kaza}

As an illustrative application of our generalized amplitude formalism we now consider scattering of an electrically charged fermion with charge $e$ off a massive magnetic monopole with magnetic charge $g$ (with  $q=eg$), reproducing the well known results of ref.~\cite{Kazama:1976fm}. In this section we eamine the lowest partial wave process, ($J=|q|-\frac{1}{2}$), and derive the celebrated helicity flip amplitude. In section~\ref{kazamahigher} we  apply our formalism to higher partial wave processes.

\subsection{Massive Fermion}
It is convenient to start with a massive Dirac fermion denoted by
\begin{eqnarray}
\psi~=~\colvec{f\\\bar{f}^{\,\dagger}}\, ,
\end{eqnarray}
where $f,\,\bar{f}$ are both LH Weyl fermions with opposite charges $e$ and $-e$. 

The $J=|q|-\frac{1}{2}$ Clebsch-Gordan coefficient for the in state can be obtained by taking $s_1\equiv s_f=1/2,\,s_2\equiv s_M=0$ and $s_3\equiv s_J=J=|q|-1/2$ in Eq.~(\ref{eq:3mas}).  That means that $\hat{s}=|q|$, and for $q>0$ the only valid 3-point $S$-matrix element is
 \begin{eqnarray}\label{eq:Blowclebsch1}
S^{\text{3-pt,in}}_{q>0}~&=&~a\,\bk{\mathbf{f}\,p^{\flat+}_{fM}}\,\bk{\mathbf{J}\,p^{\flat+}_{fM}}^{2|q|-1}\, .
\end{eqnarray}
As explained in the previous section there is only one $a$ coefficient, which we absorb in the reduced matrix element $\mathcal{M}^{J=|q|-1/2}$. Stripping away the $\bra{\mathbf{J}}^\alpha$ part, we find 
 \begin{eqnarray}\label{eq:Blowclebsch2}
C^{|q|-1/2; \, \text{in}}_{q>0}~&=&~\bk{\mathbf{f}\,p^{\flat+}_{fM}}\,\left(\ket{p^{\flat+}_{fM}}^{2|q|-1}\right)_{\left\{\alpha_1,\ldots,\alpha_{2|q|-1}\right\}}\, ,
\end{eqnarray}
and a similar one for the out state. Contracting the generalized Clebsch-Gordan factors for the \textit{in}- and \textit{out}-states, we find the {\it basis amplitude}\footnote{Since we aim to determine the $S$-matrix up to reduced matrix element $\mathcal{M}^{J}(p_c)$ we rescale our expression  by powers of $p_c$ to make the basis amplitude dimensionless.}
 \begin{eqnarray}\label{eq:Bferm1}
\mathcal{B}^{|q|-1/2}_{q>0}~&=&~\frac{\bk{\mathbf{f}\,p^{\flat+}_{fM}}\bk{\mathbf{f'}\,p^{\flat+}_{f'M'}}}{4p^2_c}\,\,{\left(\frac{\bk{p^{\flat+}_{fM}p^{\flat+}_{f'M'}}}{2p_c}\right)}^{2|q|-1}\,.
\end{eqnarray}
We can repeat the exercises for $q<0$, obtaining
 \begin{eqnarray}\label{eq:Bferm2}
\mathcal{B}^{|q|-1/2}_{q<0}~&=&~\frac{\bk{\mathbf{f}\,p^{\flat-}_{fM}}\bk{\mathbf{f^\prime}\,p^{\flat-}_{f'M'}}}{4p^2_c}\,{\left(\frac{\bk{p^{\flat-}_{fM}p^{\flat-}_{f'M'}}}{2p_c}\right)}^{2|q|-1}\,.
\end{eqnarray}

\subsection{The massless limit}

In the massless fermion  limit the particles are labeled by their helicity. Overall there are four possible choices, namely helicity $\pm\frac{1}{2}$ for the initial fermion (particle 1) and helicity $\pm\frac{1}{2}$ for the final fermion (particle 3).
In our all-outgoing convention, the helicity flip process involves the same helicity for the initial state and the final state fermions, while in the non-flip process they have opposite helicity.

The allowed processes for external fermions of charge $e$ are
\begin{eqnarray}
\text{Helicity non-flip}:&&~~~\,f\,+\,M~\rightarrow~\,f\,\,\,+\,M~~~,~~~\,\bar{f}^{\,\dagger}\,+\,M~\rightarrow~\bar{f}^{\,\dagger}\,+\,M\nonumber\\
\text{Helicity flip}:~~~~~&&~~~\,f\,+\,M~\rightarrow~\bar{f}^{\,\dagger}\,+\,M~~~,~~~\,\bar{f}^{\,\dagger}\,+\,M~\rightarrow~f\,~\,+\,M\,.
\label{eq:4processes}
\end{eqnarray}

We first consider the last process in Eq. (\ref{eq:4processes}), the right-handed incoming fermion (helicity $+1/2$) and the left-handed outgoing fermion (helicity $-1/2$). In the {\it out-out} formalism this corresponds to both fermions having helicity $-1/2$. We can take the massless limit of Eqs (\ref{eq:Bferm1}) and (\ref{eq:Bferm2}) by simply unbolding $\bra{\mathbf{f}},\,\bra{\mathbf{f'}}$ spinors \cite{Arkani-Hamed:2017jhn}.
 \begin{eqnarray}\label{eq:Bferm1unbpp}
\mathcal{B}^{|q|-\frac{1}{2}}~=~\frac{\bk{f\,p^{\flat\pm}_{fM}}\bk{f'\,p^{\flat\pm}_{f'M'}}}{4p^2_c}\,\,{\left(\frac{\bk{p^{\flat\pm}_{fM}p^{\flat\pm}_{f'M'}}}{2p_c}\right)}^{2|q|-1}&\text{for}~\,\text{sgn}(q)=\pm 1\,
\end{eqnarray}
We further note that the helicity flip amplitude Eq.~(\ref{eq:Bferm1unbpp}) is only non-trivial for $q<0$. Indeed, in the $m_i\rightarrow 0 $ limit the spinor $\ket{p^{\flat+}_{ij}}$ is parallel to $\ket{i}$ and,  according to Eq.~(\ref{eq:COMforms}), $\bk{f\,p^{\flat+}_{fM}}=\bk{f'\,p^{\flat+}_{f'M'}}=0$. The  vanishing of the $S$-matrix element for $q>0$ has a simple intuitive physical explanation. When $q>0$ the EM field component of the magnetically modified angular momentum operator (\ref{eq:magneticL}) points towards the monopole and has eigenvalues $q, q+1, q+2, \ldots$ Since we are considering the right-handed incoming fermion the minimal value of the $z$-component of the total angular momentum will be $q+1/2$ which is not part of the lowest partial wave state corresponding to $J= |q|-1/2$. One can similarly see that the outgoing left-handed particle can not be a part of the lowest partial wave when $q>0$.

Similarly, let us consider the helicity-flip amplitude where the incoming fermion is left-handed while the outgoing fermion is right-handed. In the {\it out-out} formalism this corresponds to both massless fermions having helicity $+\frac{1}{2}$. In this case we can't simply \textit{unbold} the $\bra{\mathbf{f}},\,\bra{\mathbf{f'}}$ spinors, but instead have to replace them with the Parity-conjugates\footnote{We use the properly normalized $\bra{\hat{\eta}_i}$ instead of $\bra{\eta_i}=m_i\bra{\hat{\eta}_i}$ and absorb the normalization in our reduced matrix element.} of $\bra{f}$ and $\bra{f'}$, denoted by $\bra{\hat{\eta}_f},\,\bra{\hat{\eta}_{f'}}$,
 \begin{eqnarray}\label{eq:Bferm1unbmm}
\mathcal{B}^{|q|-\frac{1}{2}}~=~\frac{\bk{\hat{\eta}_f\,p^{\flat\pm}_{fM}}\bk{\hat{\eta}_{f'}\,p^{\flat\pm}_{f'M'}}}{4p^2_c}\,\,{\left(\frac{\bk{p^{\flat\pm}_{fM}p^{\flat\pm}_{f'M'}}}{2p_c}\right)}^{2|q|-1}\,\text{for}~\,\text{sgn}(q)=\pm 1\,
\end{eqnarray}
This time, Eq.~(\ref{eq:COMforms}) tells us that $\bk{\hat{\eta}_f\,p^{\flat-}_{fM}}=\bk{\hat{\eta}_{f'}\,p^{\flat-}_{f'M'}}=0$, and so the $S$-matrix vanishes for $q<0$. Once again, there is a simple physical explanation of this fact: neither a left-handed incoming particle nor a right-handed outgoing particle can be a be part of the $J=|q|-\frac{1}{2}$ partial wave when $q<0$. %
Therefore, we find that the only non-vanishing amplitude basis for the helicity-flip process is given by
 \begin{eqnarray}\label{eq:Bferm1unbmmnonz}
&& \mathcal{B}^{|q|-\frac{1}{2}}_{q<0}~=~\frac{\bk{f\,p^{\flat-}_{fM}}\bk{f'\,p^{\flat-}_{f'M'}}}{4p^2_c}\,\,{\left(\frac{\bk{p^{\flat-}_{fM}p^{\flat-}_{f'M'}}}{2p_c}\right)}^{2|q|-1} 
\\
&& \mathcal{B}^{|q|-\frac{1}{2}}_{q>0}~\sim~\frac{\bks{f\,p^{\flat-}_{fM}}\bks{f'\,p^{\flat-}_{f'M'}}}{4p^2_c}\,{\left(\frac{\bk{p^{\flat+}_{fM}p^{\flat+}_{f'M'}}}{2p_c}\right)}^{2|q|-1}\,
\end{eqnarray}
where once again we used Eq.~(\ref{eq:COMforms}).

One can similarly show that, regardless of the sign of $q$, the $S$-matrix element vanishes for the two remaining helicity choices:  $\left(\pm\frac{1}{2},\mp\frac{1}{2}\right)$. Mathematically, this is the consequence of the fact that now the amplitude basis is proportional to a factor of the form $\bk{f\,p^{\flat\pm}_{fM}}\bk{\hat{\eta}_{f'}\,p^{\flat\pm}_{f'M'}}$, and this vanishes for either choice of $\text{sgn} (q)$. Physically, this happens because for the helicity-non-flip process either incoming or outgoing fermion can not be a part of the lowest partial wave. In other words, at the lowest partial wave helicity-non-flip process can not occur. 

Using the explicit expressions for the helicity variables in the COM frame obtained in appendix~\ref{sec:explicit_fermion_monopole_var} we can finally write the $S$-matrix in terms of the scattering angle $\theta$. The only non-vanishing $S$-matrix element is  
\begin{eqnarray}\label{eq:aflips}
S^{|q|-\frac{1}{2}}_{f\rightarrow \bar{f}^{\,\dagger}}~&=&~\mathcal{N}~2\,|q|\,\mathcal{M}^{|q|-\frac{1}{2}}_{-\frac{1}{2},\frac{1}{2}}\,{\left[\sin\left(\frac{\theta}{2}\right)\right]}^{2|q|-1}\,~~~\text{for~} q>0\nonumber\\
S^{|q|-\frac{1}{2}}_{\bar{f}^{\,\dagger}\rightarrow f}~&=&~\mathcal{N}~2\,|q|\,\mathcal{M}^{|q|-\frac{1}{2}}_{\frac{1}{2},-\frac{1}{2}}\,{\left[\sin\left(\frac{\theta}{2}\right)\right]}^{2|q|-1}\,~~~\text{for~}q<0\,,
 \end{eqnarray}
where we have explicitly included the normalization coefficient $\mathcal{N}\equiv\sqrt{8\pi s}\,$ and the reduced matrix element $\mathcal{M}^{|q|-\frac{1}{2}}_{\mp\frac{1}{2},\pm\frac{1}{2}}$, which is \textit{angle independent}. The factor $2 |q|$ is from the prefactor $(2J+1)$ (for $J=|q|-1/2$) introduced in the definition of the $S$-matrix Eq.~(\ref{eq:Def_of_S_matrix_CG_M}). 
Note that for future convenience we have used the $\textit{in-out}$ notation for the physical helicities of incoming and outgoing fermions denoted as the subscripts $\mathcal{M}_{-h_{\rm in},h_{\rm out}}$, where $h_{\rm in}, h_{\rm out}$ are helicities in {\it out-out} formalism. In general, one needs a dynamical input to determine $\mathcal{M}$ in Eq.~(\ref{eq:aflips}). However, as we will show in section~\ref{kazamahigher} the higher partial waves do not contribute to the helicity-flip matrix element. When combined with the unitarity conditions (see section~\ref{sec:unit} for a detailed discussion) this implies that
\begin{eqnarray}\label{eq:Mlowest}
\left|\mathcal{M}^{|q|-\frac{1}{2}}_{-\frac{1}{2},\frac{1}{2}}\right|~=~\left|\mathcal{M}^{|q|-\frac{1}{2}}_{\frac{1}{2},-\frac{1}{2}}\right|~=~1\, .
 \end{eqnarray}
Since the two helicity-flip processes never occur at the same time (they do or do not happen   depending on the sign of $q$), we can set them to $\mp 1$. As shown in detail in appendix~\ref{sec:kazama-compare}, the lowest partial wave $S$-matrix Eq.~(\ref{eq:aflips}) with the reduced matrix elements Eq.~(\ref{eq:Mlowest}) exactly reproduces the QM calculation of \cite{Kazama:1976fm}. 

The result is rather interesting: in the limit of massless fermions, the $S$-matrix element is only non-vanishing for processes where the products of fermion helicities, $h_f$ and $h_{f^\prime}$, with $q$ are positive, $h_f\cdot q=h_{f'}\cdot q>0$ (in the out-out sense).  It's even more striking once we remember that this discussion is in the \text{all-outgoing} convention, and so the physical interpretation in terms of in-out states is of a \textit{positive} helicity fermion scattering into a \textit{negative} helicity fermion for $q<0$, or of a \textit{negative} helicity fermion scattering into a \textit{positive} helicity fermion for $q>0$. In other words, our electric-magnetic $S$-matrix has a \textit{selection rule} that tells us that the lowest partial wave always involves a helicity flip! In particular, forward or elastic scattering is forbidden by our selection rule since it does not flip the helicity of the fermion.
This is the well-known Kazama-Yang result~\cite{Kazama:1976fm}, and also the precursor of the Rubakov-Callan effect \cite{Rubakov:1981rg,Callan:1982ac} in the scattering of two fermions and a monopole. 

\section{Fermion-Monopole Scattering: Higher Partial Waves}\label{kazamahigher}
\subsection{Massive fermions}
We now consider the $S$-matrix elements for the higher partial waves in the fermion-monopole scattering process. Once again, it is convenient to start with a massive fermion. Following our derivation of the generalized Clebsch-Gordan coefficients, we have\footnote{Notice that this result is valid for all $J$, including the lowest partial wave case $J=|q|-1/2$.}
\begin{eqnarray}\label{eq:Bferm}
\mathcal{B}^{J}~\sim\sum_\sigma\sum_{\sigma^\prime} a_{\sigma}a_{\sigma^\prime}^\prime\frac{\bk{\mathbf{f}\,p^{\flat \sigma}_{fM}}\bk{\mathbf{f'}\,p^{\flat \sigma^\prime}_{f'M'}}}{4p^2_c}\,\tilde{\mathcal{B}}^J({- q_\sigma,-q_{\sigma^\prime})}\,,
\end{eqnarray}
where sum is taken over $\sigma=(+,-)$, $\sigma^\prime=(+,-)$, while  $q_+=q-\frac{1}{2}$, $q_-=q+\frac{1}{2}$. We also included the coefficients $a_{\sigma}\,(a_{\sigma}^\prime)$ for the two possible tensor structures in the in (out) 3-point $S$-matrix elements.
The $\tilde{\mathcal{B}}^J$ are given by
 \begin{eqnarray}\label{eq:BfacJ}
\tilde{\mathcal{B}}^J({\Delta,\Delta'})~&=&~\frac{1}{(2p_c)^{2J}}\,{\left(\bra{p^{\flat-}_{fM}}^{J+\Delta}\bra{p^{\flat+}_{fM}}^{J-\Delta}\right)}^{\left\{\alpha_1,\ldots,\alpha_{2J}\right\}}\,{\left(\ket{p^{\flat-}_{f'M'}}^{J+\Delta'}\ket{p^{\flat+}_{f'M'}}^{J-\Delta'}\right)}_{\left\{\alpha_1,\ldots,\alpha_{2J}\right\}}\,.\nonumber\\
\end{eqnarray}
Using Eq.~(\ref{eq:BWig}) from appendix~\ref{sec:apWig}, in the COM frame these become
 \begin{eqnarray}\label{eq:BfacJfromap}
\tilde{\mathcal{B}}^J (\Delta,\Delta')~&=&~(-1)^{J-\Delta'}\,\mathcal{D}^{J*}_{-\Delta,\Delta'}\left(\Omega_c\right)\,.
\end{eqnarray}
where $\Omega_c=\{\theta_c,\phi_c\}$ is the direction of the outgoing COM momenta (we chose the COM frame such that $\phi_c=0$). Here $\mathcal{D}^J_{\Delta ,-\Delta'}(\Omega)$ is the Wigner matrix \cite{Wigner:1939cj,Varshalovich:1988ye}
 \begin{eqnarray}\label{eq:Wm}
\mathcal{D}^J_{-\Delta ,\Delta'}(\Omega)~\equiv~\mathcal{D}^J_{-\Delta ,\Delta'}(\phi,\theta,-\phi)~=~e^{i\phi(\Delta+\Delta')}\,d^J_{-\Delta ,\Delta'}(\theta)\,.
\end{eqnarray}
The standard definition of the Wigner d-matrix is $d^J_{m, m'}(\theta)=\bk{J,m|\exp(-i \theta J_y)|J,m'}$. The emergence of these specific $\mathcal{D}$-matrices is particularly satisfying, because they also go by another name: the spin-weighted spherical harmonics ${}_qY_{l,m}$ \cite{Wu:1976ge,Schwinger:1976fr}, or monopole harmonics~\cite{Wu:1976ge,Kazama:1976fm}. Specifically\footnote{Our ${}_qY_{lm}$ are defined according to the b-hemisphere definition of  \cite{Wu:1976ge}}:
 \begin{eqnarray}\label{eq:DYqlmb}
\mathcal{D}^{l*}_{q,m}\left(\Omega\right)~=~\sqrt{\frac{4\pi}{2l+1}}\,\,{}_qY_{l,m}\left(-\Omega\right)\, ,
\end{eqnarray}
where $-\Omega=\left(\pi-\theta,-\phi\right)$. Monopole harmonics emerge in the solution of the Klein-Gordon or Dirac equations in the presence of a background magnetic field of a monopole \cite{Wu:1976ge,Kazama:1976fm,Boulware:1976tv}. It is reassuring to see them arise here in a completely relativistic setting, and based solely on LG and angular momentum arguments. 

The $J$-partial wave matrix element for the COM scattering of a massive scalar monopole and a massive fermion is then
 \begin{eqnarray}\label{eq:massiveKaz}
 &&S^J~=~\mathcal{N}\,\,(2J+1)\,\,\frac{\mathcal{M}^{J}}{4\,p^2_c}\nonumber\\
 &&\left\{a_1a'_1\,\bk{\mathbf{f}\,p^{\flat-}_{fM}}\,\bk{\mathbf{f'}\,p^{\flat-}_{f'M'}}\,\mathcal{D}^{J*}_{q+\frac{1}{2},-q-\frac{1}{2}}\left(\Omega_c\right)\,+\,a_2a'_1\,\bk{\mathbf{f}\,p^{\flat+}_{fM}}\,\bk{\mathbf{f'}\,p^{\flat-}_{f'M'}}\,\mathcal{D}^{J*}_{q-\frac{1}{2},-q-\frac{1}{2}}\left(\Omega_c\right)\,\right.\nonumber\\
 &&\left.~\,\,a_1a'_2\,\bk{\mathbf{f}\,p^{\flat-}_{fM}}\bk{\mathbf{f'}\,p^{\flat+}_{f'M'}}\,\,\mathcal{D}^{J*}_{q+\frac{1}{2},-q+\frac{1}{2}}\left(\Omega_c\right)\,+\,a_2a'_2\,\bk{\mathbf{f}\,p^{\flat+}_{fM}}\bk{\mathbf{f'}\,p^{\flat+}_{f'M'}}\,\,\mathcal{D}^{J*}_{q-\frac{1}{2},-q+\frac{1}{2}}\left(\Omega_c\right)\right\}\, ,\nonumber\\
\end{eqnarray}
where the $(-1)^{J-\Delta'}$ prefactors have been absorbed into the coefficients $a'_i$, and $\mathcal{N}\equiv\sqrt{8\pi s}\,$.

\subsection{Massless fermion}\label{sec:highmassless}

We now consider the massless limit for the fermions in the $J>|q|-\frac{1}{2}$ partial waves. The $S$-matrix Eq.~(\ref{eq:massiveKaz}) contains all of the possible
helicity assignments, and so we can immediately extract the individual helicity amplitudes. For instance, the $S$-matrix for a helicity non-flip process $f \to f$ is obtained by unbolding the finial state massive fermion variable, and replacing the initial massive variable with $P$-conjugate $\hat{\eta}$-variable. Under this replacements, only the second term survives and Eq.~(\ref{eq:massiveKaz}) simplifies significantly to
 \begin{eqnarray}
 &&S^J_{f\rightarrow f}~=~\mathcal{N}\,\,(2J+1)\,\,\mathcal{M}^{J}_{\frac{1}{2},-\frac{1}{2}}\,\,\mathcal{D}^{J*}_{q-\frac{1}{2},-q-\frac{1}{2}}\left(\Omega_c\right)\, ,
\end{eqnarray}
where we dropped the $\frac{\bks{f\,p^{\flat-}_{f M}}\bk{f\,p^{\flat-}_{f M}}}{4p^2_c}$ factor, which equals to $1$ in the COM frame. Other cases can be worked out easily, and the general results are summarized in a compact expression as
 \begin{eqnarray}\label{eq:kazj}
 &&S^J_{h_{\text{in}}\rightarrow h_\text{out}}~=~\mathcal{N}\,\,(2J+1)\,\,\mathcal{M}^{J}_{-h_{\text{in}},h_{\text{out}}}\,\,\mathcal{D}^{J*}_{q-h_{\text{in}},-q+h_{\text{out}}}\left(\Omega_c\right)\, .
\end{eqnarray}
As shown in appendix~\ref{sec:kazama-compare}, Eq.~(\ref{eq:kazj}) exactly reproduces the angular dependence of the higher partial wave amplitudes in \cite{Kazama:1976fm}, obtained by a brute force solution of the Dirac equation in a monopole background.\footnote{We remind the reader that $h_{\text{in}},\,h_{\text{out}}$ are defined in the \textit{all-outgoing} convention, and so an incoming $f\,\left(\bar{f}^{\,\dagger}\right)$ has helicity $h_{\text{in}}=\frac{1}{2}\,\left(-\frac{1}{2}\right)$, while an outgoing $f\,\left(\bar{f}^{\,\dagger}\right)$ has helicity $h_{\text{out}}=-\frac{1}{2}\,\left(\frac{1}{2}\right)$. Note also that the indices on $\mathcal{M}^{J}$ are $-h_{\text{in}}$ and $h_{\text{out}}$, such that the labeling of $\mathcal{M}^{J}$ respects particle kind ($f$ or $\bar{f}^{\,\dagger}$) rather than helicity in the $\textit{out-out}$ convention: $-\frac{1}{2} \to f$ and $+\frac{1}{2} \to \bar{f}^\dagger$. This will be useful to keep in mind when considering $\mathcal{M}^{J\dagger}$.}

As in  textbook QM scattering in a central potential, our partial wave expansion only determines the angular dependence of each partial wave, while the relative magnitude of the different partial waves is determined dynamically in the form of \textit{phase shifts}. For the lowest partial wave, our selection rule forbids forward scattering, and so the full partial amplitude was completely fixed by unitarity. In contrast, for the higher partial waves, unitarity alone does not uniquely determine the amplitude, and some knowledge of the underlying dynamics is needed to specify the reduced matrix elements. To this end we extract the reduced matrix elements for the \textit{helicity non-flip} amplitude from \cite{Kazama:1976fm}:
\begin{equation}\label{eq:mjKazama}
\mathcal{M}^J_{\pm\frac{1}{2},\pm\frac{1}{2}}=e^{-i\pi \mu},
\end{equation}
where $\mu=\sqrt{\left(J+\frac{1}{2}\right)^2-q^2}$. One can see that  these are indeed merely phase shifts, and they are the only dynamical information needed to completely fix the $S$-matrix. The unitarity condition discussed in the next section then leads to
\begin{eqnarray}\label{eq:kazjtheirs4}
\left\vert \mathcal{M}^J_{\pm\frac{1}{2},\mp\frac{1}{2}} \right\vert^2~=~1-\left\vert \mathcal{M}^J_{\pm\frac{1}{2},\pm\frac{1}{2}}\right\vert^2~=~0\, ,
\end{eqnarray}
so the helicity-flip processes for $J>|q|-\frac{1}{2}$ vanish simply because a $100\%$ of the probability goes to the helicity non-flip process Eq.~(\ref{eq:kazj}). 

To emphasize what we have achieved, note that \textit{all} of the new information gained from the full solution of the QM scattering problem can be summarized in the phase shift Eq.~(\ref{eq:mjKazama}). In this paper we reproduced everything else based on LG and partial wave decomposition alone, in a manifestly relativistic setting. In particular, we reproduced the full angular dependence of all partial waves and the selection rule that requires a helicity-flip in the lowest partial wave.

\section{Partial Wave Unitarity}\label{sec:unit}

To complete our analysis of charged fermion scattering off a massive scalar monopole, we need to discuss partial wave unitarity. Here we follow the standard derivation of partial wave unitarity given in \cite{Pilkuhn:1979ps}, generalizing it to the electric-magnetic scattering case. Unitarity of the $S$-matrix implies
 \begin{eqnarray}\label{eq:2unit}
\frac{p_c}{16\pi^2\sqrt{s}}\,\int\,d\Omega_m~\sum_{ab}\left(S_{(f M)_i\rightarrow ab}\,S^*_{(f^{\dagger}  M)_f\rightarrow a^\dagger b^{\dagger} }\right)~=~\frac{16\pi^2\sqrt{s}}{p_c}\,\delta(\Omega_c)\, ,
\end{eqnarray}
where the momenta of $f_i$ ($M_i$)  are directed along $\pm\hat{z}$ and the momenta of $f_f$ ($M_f$) are directed along $\pm\hat{\Omega}_c$ with the angles $(\theta_c,\phi_c)$. The intermediate states $a,b$ can be either $(f_m,M_m)$ or $(\bar{f}^\dagger_m,M_m)$ with their momenta along $\pm\hat{\Omega}_m$ with the angles $(\theta_m,\phi_m)$.\footnote{Currently, we assume that the complete set of possible intermediate state consists of fermion and monopole pair $\{ f, M \}$ (with all possible choices of fermion helicity). Of course, it is certainly possible to have a microscopic theory containing other possible states, e.g.~dyon pair, or multi-particle states. However, note that what the $S$-matrix method does is to provide $S$-matrices consistent with the assumption of spectrum. Indeed, under this assumption, we find results in complete agreement with the full QM calculation with the {\it same} assumption made here.}
We now wish to perform a partial wave expansion of the unitarity relation~(\ref{eq:2unit}), in order to obtain a partial wave unitarity condition for our $S$-matrix. We begin by expanding the relevant $S$-matrix elements in partial waves, using Eq.~(\ref{eq:kazj}), which we repeat here for completeness:
 \begin{eqnarray}\label{eq:kazj2}
 &&S_{h_{\text{in}}\rightarrow h_\text{out}}~=~\mathcal{N}\,\sum_J\,(2J+1)\,\,\mathcal{M}^{J}_{-h_{\text{in}},h_{\text{out}}}\,\,\mathcal{D}^{J*}_{q-h_{\text{in}},-q+h_{\text{out}}}\left(\Omega_m\right)\, ,
\end{eqnarray}
where $\mathcal{N}\equiv\sqrt{8\pi s}\,$ is our usual normalization factor. Note that here, in contrast with the original Eq.~(\ref{eq:kazj}), the argument of the $\mathcal{D}$-matrix is $\Omega_m$ rather than $\Omega_c$. This is because we are considering the $S$-matrix for an in-state with COM momenta along the $\hat{z}$ axis and an out-state along the $\pm\hat{\Omega}_m$ direction. Similarly, we expand the inverse process as
 \begin{eqnarray}\label{eq:kazj3}
 &&S_{h_{\text{in}}\rightarrow h_\text{out}}~=~\mathcal{N}\,\sum_J\,(2J+1)\,\,\mathcal{M}^{J}_{-h_{\text{in}},h_{\text{out}}}\,\,\sum_{p=-J}^J\,\mathcal{D}^{J}_{p,q-h_{\text{in}}}\left(\Omega_c\right)\,\mathcal{D}^{J*}_{p,-q+h_{\text{out}}}\left(\Omega_m\right)\, .
\end{eqnarray}
This time we need two $\mathcal{D}$-matrices because we start from an in-state in the direction $\pm\hat{\Omega}_c$ and go to an out-state along $\pm\hat{\Omega}_m$. The explicit derivation of this particular angular dependence is presented in appendix~\ref{sec:apWig}. Substituting the above expansions in Eq.~(\ref{eq:2unit}), the unitarity relation becomes
 \begin{eqnarray}\label{eq:2units1}
 &&\frac{1}{16\pi^2}\,\int\,d\Omega_m~\sum_{J,J'}\,(2J+1)\,(2J'+1)\,\cdot\nonumber\\
&&\left\{~~\mathcal{M}^{J}_{-\frac{1}{2},-\frac{1}{2}}\,\mathcal{M}^{J'\dagger}_{-\frac{1}{2},-\frac{1}{2}}\,\,\,\mathcal{D}^{J*}_{q-\frac{1}{2},-q-\frac{1}{2}}\left(\Omega_m\right)\sum_{p=-J'}^{J'}\,\mathcal{D}^{J'*}_{p,q+\frac{1}{2}}\left(\Omega_c\right)\,\mathcal{D}^{J'}_{p,-q-\frac{1}{2}}\left(\Omega_m\right)\,\right.\nonumber\\
&&~~\left.+\,\mathcal{M}^{J}_{-\frac{1}{2},~\frac{1}{2}}\,\mathcal{M}^{J'\dagger}_{~\frac{1}{2},-\frac{1}{2}}\,\,\,\mathcal{D}^{J*}_{q-\frac{1}{2},-q+\frac{1}{2}}\left(\Omega_m\right)\sum_{p=-J'}^{J'}\,\mathcal{D}^{J'*}_{p,q+\frac{1}{2}}\left(\Omega_c\right)\,\mathcal{D}^{J'}_{p,-q+\frac{1}{2}}\left(\Omega_m\right)\right\}~=~\delta(\Omega_c)\, .\nonumber\\
\end{eqnarray}
We can perform the $\Omega_m$ integration using the orthogonality condition for $\mathcal{D}^{J}_{m,b}\left(\Omega_m\right)$, 
 \begin{eqnarray}\label{eq:Dido}
\int\,d\Omega_m~\mathcal{D}^{J*}_{a,b}\left(\Omega_m\right)\,\mathcal{D}^{J'}_{a',b'}\left(\Omega_m\right)~=~\frac{4\pi}{2J+1}\,\delta_{aa'}\,\delta_{bb'}\,\delta_{JJ'}\, .
\end{eqnarray}
Using this relation, our expression simplifies to
 \begin{eqnarray}\label{eq:2units3}
 &&\frac{1}{4\pi}\,\sum_{J}\,(2J+1)\,\left(\mathcal{M}^{J}\mathcal{M}^{J\dagger}\right)_{-\frac{1}{2},-\frac{1}{2}}\,\mathcal{D}^{J*}_{q-\frac{1}{2},q+\frac{1}{2}}\left(\Omega_c\right)~=~\delta(\Omega_c)\, .
\end{eqnarray}
Eq.~(\ref{eq:2units3}) is the unitarity relation applied to $f+M\rightarrow f+M$ scattering. Repeating the same steps for $f,\,\bar{f}^{\,\dagger}$ in the in and out state, we get the general relation
Repeating this derivation for all other in/out- states, we get
 \begin{eqnarray}\label{eq:2units4}
 &&\frac{1}{4\pi}\,\sum_{J}\,(2J+1)\,\left(\mathcal{M}^{J}\mathcal{M}^{J\dagger}\right)_{-h_{\text{in}},h_{\text{out}}}\,\mathcal{D}^{J*}_{q-h_{\text{in}},q-h_{\text{out}}}\left(\Omega_c\right)~=~\delta_{-h_{\text{in}},h_{\text{out}}}\,\delta(\Omega_c)\, .
\end{eqnarray}
Multiplying by $\mathcal{D}^{J}_{q-h_{\text{in}},q-h_{\text{out}}}\left(\Omega_c\right)$ and using Eq.~(\ref{eq:Dido}), we have 
 \begin{eqnarray}\label{eq:unitrel}
 &&\mathcal{M}^{J}\mathcal{M}^{J\dagger}~=~I\, ,
\end{eqnarray}
where $\mathcal{M}^{J}$ is the $2\times2$ matrix representing $f$ or $\bar{f}^{\,\dagger}$ in the in / out state, and $I$ is the $2\times2$ identity matrix. In other words, the unitarity of the $S$-matrix leads to the unitarity of each individual \textit{reduced matrix element} $\mathcal{M}^{J}$. This is also the standard result for non-magnetic amplitudes \cite{Pilkuhn:1979ps}, which leads to the partial-wave unitarity bound \cite{Griest:1989wd}. Here we see that it holds for the electric-magnetic case as well, even though the eigenfunctions of the partial wave decomposition are modified by the extra angular momentum in the EM field. The unitarity condition Eq.~(\ref{eq:unitrel}) is key in reproducing the full helicity-flip amplitude for the $J=|q|-\frac{1}{2}$ partial wave in section~\ref{sec:kaza}, as well as the vanishing of the helicity-flip amplitudes for $J>|q|-\frac{1}{2}$ in section~\ref{sec:highmassless} (assuming that the helicity non-flip process is given by Eq.~(\ref{eq:mjKazama})).

\section{Conclusions}

In this paper we have initiated the systematic study of electric-magnetic scattering amplitudes, using on-shell methods. We have identified the multi-particle representations of the Poincar\'e group that are necessary to incorporate asymptotic states with both electric and magnetic charges. At the heart of our study is the appearance of a new pairwise LG and its corresponding pairwise helicity, which describe the transformation of electric-magnetic multi-particle states relative to the direct product of the one-particle states. This pairwise helicity is non-zero for a charge-monopole pair and corresponds to the angular momentum stored in the asymptotic electromagnetic field, which is appropriately quantized if  Dirac-Schwinger-Zwanziger charge quantization is satisfied. This novel pairwise helicity gains a simple and intuitive implementation in the scattering amplitude formalism, through the definition of pairwise spinor-helicity variables. We then used the pairwise spinor-helicity variables to formulate the general rules for building the electric-magnetic $S$-matrix.  In particular, we were able to classify all 3-particle magnetic $S$-matrix elements, corresponding to decays of magnetically charged particles. Many of these electric-magnetic $S$-matrix elements are subject to simple selection rules among the spins/helicities and pairwise helicities of the various particles. In addition, we performed a pairwise LG covariant partial wave expansion for the generic $2\to 2$ fermion-monopole scattering amplitude. For the lowest partial wave, our LG based selection rules allowed us to derive the famous helicity flip for the lowest partial wave. Furthermore, the well-known monopole spherical-harmonics appear naturally in our formalism, and the general results of~\cite{Kazama:1976fm} are fully reproduced up to dynamics-dependent phase shifts. We never have to introduce a Dirac string, and the resulting $S$-matrix elements are always manifestly Lorentz invariant. For monopoles that do not satisfy Dirac-Schwinger-Zwanziger charge quantization due to kinetic mixing with a hidden sector photon \cite{Terning:2018lsv} a separate treatment is needed \cite{Terningprogress}. 

Recently the authors of ref.~\cite{Hannesdottir:2019opa}  discussed  the need for a more careful definition of the $S$-matrix; they define a ``hard'' $S$-matrix by evolving the asymptotic states with an asymptotic Hamiltonian
which is not the free Hamiltonian, but allows for the emission and absorption of massless photons.  This evolution builds up a cloud of photons representing the Coulomb fields of the charged in and out particles.  In the presence of both electric and magnetic charges the Coulomb fields carry additional angular momentum which we have included explicitly using the pairwise LG.  It would be interesting to see how this angular momentum could be handled in the ``hard'' $S$-matrix formalism. It will also be interesting to consider the double copy relation between dyons and Taub-NUT spaces \cite{Caron-Huot:2018ape,Moynihan:2020gxj,Kol:2020ucd} in light of our results.

\section*{Acknowledgments}
We thank Curt Callan, Hsin-Chia Cheng, Kit Colwell, Markus Luty, Juan Maldacena, Joe Polchinski, David Tong, and Chris Verhaaren for useful discussions and comments over the ten years that we have been working toward this paper.  
We thank the Aspen Center for Physics where part of this work was completed.  C.C. and J.T. thank the Kavli Institute for Theoretical Physics where part of this work was completed. J.T. thanks CERN where part of this work was completed.  The research of C.C. has been supported in part by the NSF grant PHY-2014071  and in part by a U.S.-Israeli BSF grant. 
S.H. is supported by the NSF grant PHY-2014071, and by Cornell University through the Hans Bethe Postdoctoral Fellowship.
Y.S. and M.W. are supported in part by the NSF grant  PHY-1915005.
J.T. is supported by the DOE under grant  DE-SC-0009999. OT is supported in part by the DOE under grant DE-AC02-05CH11231.

\appendix

\section{Notation}  
We work in mostly-minus signature $(+,-,-,-)$. Our Pauli matrices are defined as
 \begin{eqnarray}\label{eq:psap1}
\left( \sigma^\mu \right)_{\alpha\dot{\alpha}}~=~\left(I\,,\,\vec{\sigma}\right)~~,~~\left( \bar{\sigma}^{\mu} \right)^{\dot{\alpha}\alpha}~=~\left(I\,,\,-\vec{\sigma}\right)\,,
 \end{eqnarray}
where
 \begin{eqnarray}
\sigma^1~=~\colmatt{0&1\\1&0}~,~\sigma^2~=~\colmatt{0&-i\\i&0}~,~\sigma^3~=~\colmatt{1&0\\0&-1}\, .
 \end{eqnarray}
Undotted indices are raised and lowered by the two index epsilon symbol
 \begin{eqnarray}
\epsilon^{\alpha\beta}~=~\epsilon_{\alpha\beta}~=~\colmatt{0&1\\-1&0}\, ,
 \end{eqnarray}
following a northwest-southeast convention:
 \begin{eqnarray}
\lambda^\alpha~=~\epsilon^{\alpha\beta}\lambda_\beta~,~\lambda_\alpha~=~\lambda^\beta \epsilon_{\beta\alpha}\, .
 \end{eqnarray}
Similarly, dotted indices are raised and lowered with 
 \begin{eqnarray}
\epsilon^{\dot{\alpha}\dot{\beta}}~=~\epsilon_{\dot{\alpha}\dot{\beta}}~=~\colmatt{0&-1\\1&0}\, ,
 \end{eqnarray}
following a northwest-southeast convention:
 \begin{eqnarray}
\tilde{\lambda}^{\dot{\alpha}}~=~\tilde{\lambda}_{\dot{\beta}}\,\epsilon^{\dot{\beta}\dot{\alpha}}
~,~\tilde{\lambda}_{\dot{\alpha}}~=~\epsilon_{\dot{\alpha}\dot{\beta}}\,\tilde{\lambda}^{\dot{\beta}}\, .
 \end{eqnarray}

We define symmetrized products as:
 \begin{eqnarray}
&&\left(~\ket{a_1}^{n_1}\,\cdot\ldots\cdot\ket{a_k}^{n_k}~\right)_{\left\{\alpha_1,\ldots\alpha_{2J}\right\}}~\equiv~\nonumber\\
&&\mathcal{N}~\sum_{\sigma_k}\,\ket{a_1}_{\alpha_{\sigma_k(1)}}\cdot\ldots\cdot\ket{a_1}_{\alpha_{\sigma_k(n_1)}}\cdot\ldots\cdot\ket{a_{k}}_{\alpha_{\sigma_k(2J-n_k+1)}}\cdot\ldots\cdot\ket{a_k}_{\alpha_{\sigma_k(2J)}}\, ,\nonumber\\
 \end{eqnarray}
where $\sum\,n_i\,=\,2J$, and the sum is over permutations on $k$ elements. We choose the normalization factor to be
 \begin{eqnarray}\label{eq:normsym}
\mathcal{N}~=~\left[(2J)!\,\prod_{i=1}^k\,(n_i)!\right]^{-\frac{1}{2}}\, .
 \end{eqnarray}
This choice of normalization  gives us Wigner $\mathcal{D}$-matrices when contracting symmetric products of spinors in the COM frame.
\subsection{Conventions}

The fermions in our paper are all left-handed Weyl, while their hermitian conjugates are right-handed:
 \begin{eqnarray}\label{eq:ferms}
f~\equiv~\text{LH Weyl}~~~~,~~~~f^\dagger~\equiv~\text{RH Weyl}\, .
 \end{eqnarray}

We work in the all-outgoing convention for the $S$-matrix, for consistency with the rest of the scattering $S$-matrix literature. In practice it means that $h=\frac{1}{2}\,(-\frac{1}{2})$ for the initial (i.e.~originally incoming but crossed to outgoing) LH (RH) Weyl fermions, and $h=-\frac{1}{2}\,(\frac{1}{2})$ for the final (outgoing) LH (RH) Weyl fermions. 

Reduced matrix elements are labeled as
\begin{eqnarray}\label{eq:Mlab}
\mathcal{M}^{J}_{-h_{\text{in}},h_{\text{out}}}\, 
\end{eqnarray}
in our all-outgoing convention, $h_{\text{in}}=\frac{1}{2}\,(-\frac{1}{2})$ for incoming $f\,(\bar{f}^{\,\dagger})$, and $h_{\text{out}}=-\frac{1}{2}\,(\frac{1}{2})$ for outgoing $f\,(\bar{f}^{\,\dagger})$. This means that the labels on $\mathcal{M}^J$ respect particle identity: $\mathcal{M}^J_{-\frac{1}{2},\frac{1}{2}}$ is for $f\rightarrow \bar{f}^{\,\dagger}$ while $\mathcal{M}^J_{\frac{1}{2},-\frac{1}{2}}$ is for $\bar{f}^{\,\dagger}\rightarrow f$, etc.

\section{Spinor-helicity variables in the COM frame and in the heavy monopole limit}\label{sec:explicit_fermion_monopole_var}
In the COM frame of a dyon pair $i,\,j$ we have
 \begin{eqnarray}\label{eq:commom}
p^\mu_i~&=&~\left(E^c_i,~+\hat{p}_c\right)\nonumber\\
p^\mu_j~&=&~\left(E^c_j,~-\hat{p}_c\right)\, ,
\end{eqnarray}
where $\hat{p}_c$ is in the direction given by $\left\{\theta_c,\phi_c\right\}$ and
\begin{eqnarray}\label{eq:psap2}
p_c~=~\sqrt{\frac{(p_i \cdot p_j)^2 - m_i^2 m_j^2}{s}}~~,~~E^c_{i,j}=\sqrt{m^2_{i,j}+p^2_{\text{c}}}\,.
 \end{eqnarray}
In this case, the Lorentz transformation $L_p$ taking the reference momenta Eq.~(\ref{eq:refmom}) to $p_i,\,p_j$ is just a rotation
\begin{eqnarray}\label{eq:psap3}
L_p~=~R_z\left(\phi_c\right)\,R_y\left(\theta_c\right)\, .
 \end{eqnarray}
Acting with the spinor version of this transformation on the reference pairwise spinors $\ket{ k^{\flat\pm}_{ij} }_\alpha,\,\sbra{ k^{\flat\pm}_{ij}}_{\dot{\alpha}}$, etc. we get
\begin{eqnarray}
\ket{ p^{\flat\pm}_{ij} }_\alpha~=~\sqrt{2p_c}\,\ket{ \mypm \hat{p}_c }_\alpha~~,~~\sbra{ p^{\flat\pm}_{ij}}_{\dot{\alpha}}~=~\sqrt{2p_c}\,\sbra{ \mypm \,\hat{p}_c }_{\dot{\alpha}}\, ,
\end{eqnarray}
where we use ``$\text{-}$'' instead of $-$ inside the brackets for ease of reading. 
In the equation above we use the notation
\begin{eqnarray}\label{eq:normsh}
\ket{ \hat{n} }_{\alpha}~\equiv~~\,\colvec{c_n\\s_n}\,~~~&,&~~~\sbra{ \hat{n} }_{\dot{\alpha}}~\equiv~~~\left(c_n~,~s^*_n\right)\nonumber\\
\ket{\, \text{-}\hat{n} }_{\alpha}~\equiv~\colvec{-s^*_n\\\,c_n\,}~~&,&~~\sbra{\,\text{-} \hat{n} }_{\dot{\alpha}}~\equiv~\left(-s_n~,~c_n\right)\, .
\end{eqnarray}
where $s_n=e^{i\phi_n}\sin\left(\frac{\theta_n}{2}\right),\,c_n=\cos\left(\frac{\theta_n}{2}\right)$. In particular, under a parity transformation $\hat{n}\leftrightarrow -\hat{n}$, we have
\begin{eqnarray}
\ket{ \hat{n} }_{\alpha}~\leftrightarrow~-e^{i\phi_n}\,\ket{\, \text{-}\hat{n} }_{\alpha}~~~,~~\sbra{ \hat{n} }_{\dot{\alpha}}~\leftrightarrow ~-e^{-i\phi_n}\,\sbra{\, \text{-}\hat{n} }_{\dot{\alpha}}\, .
\end{eqnarray}
The expressions for $\bra{\mypm\hat{n} }^{\alpha}$ and $\sket{\mypm\hat{n} }^{\dot{\alpha}}$ are obtained by raising the spinor indices with $\epsilon^{\alpha\beta}$ and $\epsilon^{\dot{\alpha}\dot{\beta}}$, following the northwest-southeast convention for $\alpha$ and the southwest-northeast convention for $\dot{\alpha}$. Explicitly,
\begin{eqnarray}
\bra{ \hat{n} }^{\alpha}~=~\left(s_n~,~-c_n\right)\,~~&,&~~~~\sket{ \hat{n} }^{\dot{\alpha}}~~=~\colvec{s^*_n\\-c_n}\nonumber\\
\bra{ \,\text{-}\hat{n} }^{\alpha}~=~~\,\,\left(c_n~,~s^*_n\right)~~~~&,&~~~~\sket{\,\text{-} \hat{n} }^{\dot{\alpha}}~=~\,\,\colvec{c_n\\s_n}\, .
\end{eqnarray}
Also, since in the center of mass frame $\hat{p}_i=-\hat{p}_j=\hat{p}_c$\,, we automatically get the following relations in the $m_i\rightarrow 0$ limit
\begin{eqnarray}
\ket{ p^{\flat+}_{ij} }_\alpha~=~\ket{\,i \,}_\alpha\,~~~~~~~~~~&,&~~~~~\sbra{ p^{\flat+}_{ij}}_{\dot{\alpha}}~=~\sbra{\,i\,}_{\dot{\alpha}}\nonumber\\
\ket{ p^{\flat-}_{ij} }_\alpha~=~\sqrt{2p_c}\,\ket{\hat{\eta}_i}_\alpha~~\,&,&~~~~~\sbra{ p^{\flat-}_{ij}}_{\dot{\alpha}}~=~\sqrt{2p_c}\,\sbra{\hat{\eta}_i}_{\dot{\alpha}}\, ,
\end{eqnarray}
where $\ket{i }_\alpha,\,\sbra{i}_{\dot{\alpha}}$ are the standard massless spinor-helicity variables, and $\ket{\hat{\eta}_i }_\alpha,\,\sbra{\hat{\eta}_i}_{\dot{\alpha}}$ are the (dimensionless) Parity-conjugate massless spinors that appear in the massless limit of the massive spinors $\ket{\mathbf{i} }^I_\alpha,\,\sbra{\mathbf{i}}^I_{\dot{\alpha}}$ (see \cite{Arkani-Hamed:2017jhn} for their definition).
Consequently, the following contractions vanish:
\begin{eqnarray}
&&\bks{p^{\flat+}_{ij}\,i}~=~\bk{i\, p^{\flat+}_{ij}}~=~\bks{\hat{\eta}_i\, p^{\flat-}_{ij}}~=~\bk{p^{\flat-}_{ij}\,\hat{\eta}_i}~=~0\nonumber\\
&&\bks{p^{\flat-}_{ij}\,i}~=~\bk{i\, p^{\flat-}_{ij}}~=~\bks{\hat{\eta}_i\, p^{\flat+}_{ij}}~=~\bk{ p^{\flat+}_{ij}\,\hat{\eta}_i}~=~2p_c\, ,
\end{eqnarray}
since $\bks{\text{-}\hat{n}\,| \,\hat{n}}=\bk{\hat{n} \,|\,\text{-}\hat{n}}=1$. Note that the above equations are Lorentz \textit{and} LG invariant, and so hold in any other reference frame as well. 

\subsection{$2\rightarrow2$ scattering in the COM frame and Wigner $\mathcal{D}$-matrices}\label{sec:apWig}
We now explicitly present the relevant formulas for $2\rightarrow 2$ scattering in the COM frame. We take the colliding momenta to be
\begin{eqnarray}
p^\mu_i~\,=~\left(E^c_i, \,\hat{n}\,p_c\,\right)~~&,&~~~p^\mu_j\,~=~\left(E^c_j, \,-\hat{n}\,p_c\,\right)\nonumber\\
\tilde{p}^\mu_i~\,=~\left(E^c_i, \,\hat{k}\,p_c\,\right)~~&,&~~~\tilde{p}^\mu_j\,~=~\left(E^c_j, \,-\hat{k}\,p_c\,\right)\, ,
\end{eqnarray}
where $\hat{n}$ is in the $(\theta_n,\phi_n)$ direction and $\hat{k}$ is in the $(\theta_k,\phi_k)$ direction. Later we will specialize to the case $\theta_n=0$ in which the initial momenta point along the $\hat{z}$ direction. From Eq.~(\ref{eq:normsh}) we have
\begin{eqnarray}\label{eq:explicit}
\bk{\,\text{-}\hat{n}\,|\,\text{-}\hat{k}}^*~&=&~\,\,\bk{\,\hat{n}\,|\,\hat{k}}~=~s_n\,c_k\,-\,c_n\,s_k\nonumber\\
-\bk{\,\hat{n}\,|\,\text{-}\hat{k}}^*~&=&~\bk{\,\text{-}\hat{n}\,|\,\hat{k}}~=~c_n\,c_k\,+\,s^*_n\,s_k\, .
\end{eqnarray}
where $s_i=e^{i\phi_i}\sin\left(\theta_i/2\right),\,c_i=\cos\left(\theta_i/2\right)$ for $i=n,k$. We put a $|$ to separate contractions involving a ``$\text{-}$'' for ease of reading. The expression for square brackets are obtained by $\bks{ab}=\bk{ba}^*$.

When writing down $2\rightarrow 2$ electric-magnetic $S$-matrix elements, we encounter the ubiquitous spinor contraction
 \begin{eqnarray}\label{eq:BfacJapp}
\tilde{\mathcal{B}}^J(\Delta,\Delta')~&=&~{\left(\bra{\,\text{-}\hat{n}}^{J+\Delta}\bra{\,\hat{n}}^{J-\Delta}\right)}^{\left\{\alpha_1,\ldots,\alpha_{2J}\right\}}\,{\left(\ket{\,\text{-}\hat{k}}^{J+\Delta'}\ket{\,\hat{k}}^{J-\Delta'}\right)}_{\left\{\alpha_1,\ldots,\alpha_{2J}\right\}}\,.~~~~~~~~~~~~
\end{eqnarray}
By simple combinatorics, this expression simplifies to the sum
 \begin{eqnarray}\label{eq:BfacJcomb}
\tilde{\mathcal{B}}^J(\Delta,\Delta')~&=&~\sum_i\,w_i~\bk{\,\text{-}\hat{n}\,|\,\text{-}\hat{k}}^i~\bk{\,\hat{n}\,|\,\hat{k}}^{i-\Delta-\Delta'}~\bk{\,\text{-}\hat{n}\,|\,\hat{k}}^{J+\Delta-i}~\bk{\,\hat{n}\,|\,\text{-}\hat{k}}^{J+\Delta'-i}\,.~~~~~~~
\end{eqnarray}
where the sum is over $\text{max}(0,\Delta+\Delta')\leq i\leq J+ \text{min}(\Delta,\Delta')$. The coefficients $\omega_i$ are combinatoric factors denoting the number of equivalent contractions \cite{Jiang:2020sdh},
\begin{equation}\label{eq:wis}
w^{i}=\frac{\sqrt{J+\Delta) !\,(J-\Delta) !\,\left(J+\Delta^{\prime}\right) !\,\left(J-\Delta^{\prime}\right) !}}{i!\,(i-\Delta-\Delta') !\,\left(J+\Delta-i\right) ! \,i !\,\left(J+\Delta'-i\right) !}\, .
\end{equation}
Note that to get $w^i$ we have used our particular normalization for symmetrized products, Eq.~(\ref{eq:normsym}). Substituting the values Eq.~(\ref{eq:explicit}) in Eq.~(\ref{eq:BfacJapp}), one can check explicitly that the following relation holds:
 \begin{eqnarray}\label{eq:BWig1}
\tilde{\mathcal{B}}^J(\Delta,\Delta')~&=&~(-1)^{J-\Delta'}\,\sum_{p=-J}^J\,\mathcal{D}^{J}_{p,-\Delta}\left(\phi_n,\theta_n,-\phi_n\right)\,\mathcal{D}^{J*}_{p,\Delta'}\left(\phi_k,\theta_k,-\phi_k\right)\,.~~~~~~~~
\end{eqnarray}
$\mathcal{D}^{J}_{m,m'}\left(\alpha,\beta,\gamma\right)$ is the Wigner $\mathcal{D}$-matrix, defined as
 \begin{eqnarray}\label{eq:Dmat}
\mathcal{D}^{J}_{m,m'}\left(\alpha,\beta,\gamma\right)~\equiv~\langle J,m \vert \mathcal{R} (\alpha,\beta,\gamma) \vert J, m' \rangle = e^{-i \,(m\, \alpha\,+\,m' \,\gamma)}~d^{J}_{m,m'}\left(\beta\right)\, ,
\end{eqnarray}
where $\mathcal{R} (\alpha,\beta,\gamma) = e^{-i \alpha J_z} e^{-i \beta J_y} e^{-i \gamma J_z}$ is a 3-dimensional rotation operator, and therefore 
\bea
d^{J}_{m,m'}\left(\beta\right)\equiv\bk{J,m\,|\,e^{-iJ_y\beta}\,|\,J,m'}.
\eea
Since our $\mathcal{D}$-matrices always involve $\gamma=-\alpha=-\phi,\,\beta=\theta$, we use the shorthand notation
 \begin{eqnarray}\label{eq:Dmatshort}
\mathcal{D}^{J}_{m,m'}\left(\Omega\right)~\equiv~\mathcal{D}^{J}_{m,m'}\left(\phi,\theta,-\phi\right)\, ,
\end{eqnarray}
where $\Omega = \{ \theta, \phi \}$.
 In the particular case where the initial momenta are along the $\pm\hat{z}$ direction, we have $\theta_n=0$, and Eq.~(\ref{eq:BWig1}) reduces to
 \begin{eqnarray}\label{eq:BWig}
\tilde{\mathcal{B}}^J(\Delta,\Delta')~&=&~(-1)^{J-\Delta'}\,\mathcal{D}^{J*}_{-\Delta,\Delta'}\left(\Omega_k\right)\,.
\end{eqnarray}
We make use of this expression in section~\ref{sec:kaza}, where we consider $2\rightarrow 2$ electric-magnetic $S$-matrix elements in the COM frame.

\subsection{The heavy particle limit}\label{sec:heavy}
In the $m_j\rightarrow\infty$ limit, Eq.~(\ref{eq:ppmeq}) leads to very simple expressions for the spatial parts of the pairwise momenta,
\begin{eqnarray}
\vec{p}^{~\,\flat\pm}_{ij}~&=&~\pm \vec{p}_i\, .
\end{eqnarray}
Note that in this limit $p_i\sim p_c$ up to $\mathcal{O}\left(m^{-1}_j\right)$ corrections. That implies
\begin{eqnarray}
\ket{ p^{\flat\pm}_{ij} }_\alpha~=~\sqrt{2p_c}\,\ket{ \pm \hat{p}_i }_\alpha~~,~~\sbra{ p^{\flat\pm}_{ij}}_{\dot{\alpha}}~=~\sqrt{2p_c}\,\sbra{ \pm \,\hat{p}_i }_{\dot{\alpha}}\, ,
\end{eqnarray}
and we are free to use all the expressions derived throughout appendix~\ref{sec:explicit_fermion_monopole_var} for the COM frame also in any other frame with the substitution $\hat{p}_c\rightarrow\hat{p}_i$. This is correct up to $\mathcal{O}\left(m^{-1}_j\right)$ corrections.

\section{Definition of the electric-magnetic $S$-matrix}\label{sec:Smat}

In this section we define the $S$-matrix rigorously following Weinberg \cite{Weinberg:1995mt}, making changes when necessary to adapt to the electric-magnetic case. We work in the Heisenberg picture, where all of the time dependence is concentrated in the operators rather than in the quantum states. As in the standard definition of the $S$-matrix, we separate the full Hamiltonian of the system into a free and interacting part, as in Eq.~(\ref{eq:Hami}).
Note that in the case of electric-magnetic scattering, the free part $H_0$ and the full Hamiltonian $H$ have \textit{different} conserved angular momentum operators,
\begin{eqnarray}\label{eq:angmomops2}
\bks{H,\vec{J}}~=~\bks{H_0,\vec{J}_{0}}~=~0,~~~\vec{J}~\neq~\vec{J}_{0}\, .
\end{eqnarray}
 This means that the Lorentz group is represented differently on the eigenstates of $H$ and $H_0$. We'll return to this point below.
 
As a first step towards the definition of the $S$-matrix, we define the eigenstates $\ket{\alpha;\,\text{free}}$ of the non-interacting part $H_0$ such that,
\begin{eqnarray}\label{eq:eogH0}
H_0\,\ket{\alpha;\,\text{free}}~=~E_\alpha\,\ket{\alpha;\,\text{free}}\,.
\end{eqnarray}
The label $\alpha$ denotes the different eigenstates of $H_0$. Since $H_0$ is free, its eigenfunctions are just direct products (or sums of direct products) of one-particle states,
\begin{eqnarray}\label{eq:multi-particleH0}
\ket{\alpha;\,\text{free}}~=~\prod_{i\in\alpha}\,\ket{p_i;s_i;\,n_i}\,,
\end{eqnarray}
where $p_i$ and $s_i$ are the momentum and spin/helicity of each particle, and $n_i$ denotes its charges and gauge representations.

As in \cite{Weinberg:1995mt}, we define our in (out) states as eigenstates of $H$. Since the interaction $V$ vanishes asymptotically, the eigenstates of $H$ and $H_0$ coincide, and we can write 
\begin{eqnarray}\label{eq:eogH}
H\,\ket{\alpha;\,\pm\,}~=~E_\alpha\,\ket{\alpha;\,\pm\,}\,,
\end{eqnarray}
where `$+$' denotes in states and `$-$' denotes out states. In Weinberg's definition, the labels in (out) define two different eigenbases of $H$, which differ by their asymptotic forms at $t\rightarrow\pm\infty$. From this limiting relation and using $J=J_0$ (valid in his case but \textit{not} in ours), he deduces how the Lorentz group is represented on in/out- states, and more importantly, that the \textit{in}- and \textit{out}- representations are identical.

In the case of an electric-magnetic $S$-matrix, $J\neq J_0$ by the non-vanishing asymptotic value of the angular momentum in the EM field. Inspired by Zwanziger \cite{Zwanziger:1972sx}, we follow an opposite route to Weinberg, namely, we \textit{define} our in out states by their \textit{different} representations under the Lorentz group, and derive the implications for the $S$-matrix. The transformation rule that we impose on our \textit{in}- and \textit{out}- states is given in Eq.~(\ref{eq:2momtrans}), and we repeat it here for completeness,
\bea\label{eq:2momtransapp}
U(\Lambda)\, \ket{p_1, \ldots ,p_n\,;\,\pm\,}~&=&~\prod_i \mathcal{D} (W_i)~\ket{\Lambda p_1, \ldots ,\Lambda p_n\,;\,\pm\,}\,e^{\pm i\,\Sigma}\nonumber\\
U_{\text{free}}(\Lambda)\, \ket{p''_1 \ldots p''_l\,;\,\text{free}}~&=&~\prod_i \mathcal{D} (W_i) \ket{\Lambda p''_1 \ldots \Lambda p''_l\,;\,\text{free}}\, ,
\eea
where $\Sigma\equiv\sum_{i>j} q_{ij} \,\phi (p_i,p_j, \Lambda)$. We explicitly present the momenta $p_i$ of the particles involved but suppress their spin/helicity labels, which are implicit in the LG transformations $\mathcal{D} (W_i)$. The magnetic part of the transformation for in/out-states is evident in the $q_{ij}$ dependence of $\Sigma$, where $q_{ij}=e_ig_j-e_jg_i$ is the pairwise helicity of each particle pair. In section~\ref{subsec:monopole-charge_asymp_state} we prove that these transformation rules constitute a unitary representation of the Lorentz group, by explicitly constructing them through the method of induced representations. The transformation rule Eq.~(\ref{eq:2momtransapp}) is a departure from Weinberg's standard definition of the $S$-matrix, in the sense that the Lorentz group is represented \textit{differently} on \textit{in}- and \textit{out}- sates.

Having defined our in/out states in terms of their representations under Lorentz transformation, we can now take their $t\rightarrow\pm\infty$ limits to get relations similar to Weinberg's Eq. 3.1.12. In these limits, we would like to make the statement that our \textit{in}- and \textit{out}- states approach free states, since the interaction term $V$ vanishes for $t\rightarrow\pm\infty$. However, our naive expectation is hindered by the extra phases in the transformation of our in- and out-states. To compensate for that, we define our \textit{compensated} free states:
 \begin{eqnarray}\label{eq:prf}
\ket{p''_1 \ldots p''_l\,;\,\left(\text{free}\,\pm\right)\,}~\equiv~\,C_{\pm}(p''_1 \ldots p''_l)\,\ket{p''_1 \ldots p''_l\,;\,\text{free}\,}\, ,
 \end{eqnarray}
where $C_\pm$ is a ``compensator'' function of the momenta which satisfies
 \begin{eqnarray}\label{eq:comp}
C_\pm(p''_1 \ldots p''_l)~&=&~e^{\pm i\Sigma}~C_{\pm}(\Lambda p''_1 \ldots \Lambda p''_l)\nonumber\\
{|C_\pm(p''_1 \ldots p''_l)|}^2~&=&~1\, .
 \end{eqnarray}
 The compensator functions are unique up to a constant phase, and we can construct them explicitly from our pairwise spinor-helicity variables, as we demonstrate for the $2\rightarrow 2$ case in section~\ref{sec:unit}.  

Because of the compensator functions, the compensated free states have the same transformation rule as their in/out- counterparts, so they can serve as the right limits at $t\rightarrow\pm \infty$. We now make this statement in a more formal manner. Since we are working in the Heisenberg picture, we define time dependent superpositions of in, out, and free states as
 \begin{eqnarray}\label{eq:superpos}
\ket{g,t;\,\pm\,}~&=&~\exp\left(-i\,H\, t\right)\,\,\,\int\,d\alpha\,g(\alpha)\,\ket{\alpha;\,\pm\,}\nonumber\\
\ket{g,t;\,\left(\text{free}\,\pm\right)}~&=&~\exp\left(-i\,H_0\, t\right)\,\int\,d\alpha\,g(\alpha)\,\ket{\alpha;\,\left(\text{free}\,\pm\right)\,}\,.
\end{eqnarray}
Taking the $t\rightarrow\pm\infty$ limit of our in/out- superpositions, and noting that $H\rightarrow H_0$ in this limit, we get the limiting forms
 \begin{eqnarray}\label{eq:multi-particleH0limitinout}
\lim_{t\rightarrow\mp\infty}\,\ket{g,t\,;\,\pm\,}~&=&~\lim_{t\rightarrow\mp\infty}\,\ket{g,t\,;\,\left(\text{free}\,\pm\right)\,}\,.
\end{eqnarray}
A different way of stating the same relation is the formal expression
 \begin{eqnarray}\label{eq:formalexp}
\ket{\alpha;\,\pm\,}~&=&~\Omega(\mp\infty)\,\ket{\alpha;\,\left(\text{free}\,\pm\right)\,}\, ,
\end{eqnarray}
where $\Omega(t)\equiv\exp(iHt)\exp(-iH_0t)$. This relation should be understood in terms of superpositions as in Eq.~(\ref{eq:superpos}).
The $S$-matrix is defined as usual as:
  \begin{eqnarray}\label{eq:multi-particlefreeout1}
S_{\beta\alpha}~=~\bk{\beta;\,-\,|\,\alpha;\,+\,}\, ,
 \end{eqnarray}
or equivalently as 
  \begin{eqnarray}\label{eq:multi-particlefreeout2}
S_{\beta\alpha}~=~\bk{\beta;\,\left(\text{free}\,-\right)\,|S|\,\alpha;\,\left(\text{free}\,+\right)\,}\, ,
 \end{eqnarray}
where $S\equiv \Omega^\dagger(\infty)\,\Omega(-\infty)$.

 \section{Zwanziger's Vectors}
The first derivation of the LG transformation for electric-magnetic $S$-matrix elements was given by Zwanziger for $q_{ij}=1$ in a seminal paper \cite{Zwanziger:1972sx}. Beyond deriving the LG transformation similarly\footnote{The main difference between our derivation and Zwanziger's original derivation is our choice of the reference momenta $k_{i,j}$ to be the COM momenta rather than the momenta in the monopole rest frame. This makes our formalism more symmetric and suitable for the introduction of pairwise spinors.} to our section~\ref{subsec:monopole-charge_asymp_state}, Zwanziger also defined LG covariant vectors, which he used to construct manifestly LG covariant $S$-matrix elements. Unfortunately, Zwanziger's vectors were explicitly Lorentz non-invariant, as they have an explicit dependence on an arbitrary direction $\hat{n}$. This was not a major detractor from his formalism, though, since all of the $\hat{n}$ dependence canceled out when taking the absolute value squared of the $S$-matrix. Our use of pairwise spinors rather than vectors eliminates this $\hat{n}$ dependence, up to our choice of the canonical Lorentz transformation $L_p$ which takes $k_{i,j}\rightarrow p_{i,j}$. However, this is no different from the usual choice of a canonical Lorentz transformation in the standard Wigner method.
The other main detractor from using Zwanziger's pairwise vectors was the fact that they have pairwise helicity $\pm1$ rather than $\pm\frac{1}{2}$, which excludes writing down $S$-matrix elements with half integer $q$. Our formalism closes this gap, and allows us to write down pairwise LG covariant $S$-matrix elements in their most general form.
\\ \quad\\
In this appendix we define Zwanziger's vectors in terms of our pairwise spinor-helicity variables, and reproduce his results from section V of \cite{Zwanziger:1972sx}.
To define LG covariant vectors, we first pick a reference vector $n^\mu$ and define:
\begin{eqnarray}\label{eq:azw}
a^\mu_+~&=&~~~~~i\,\sqrt{\frac{\bmk{p^{\flat+}_{ij}}{n}{p^{\flat-}_{ij}}}{\bmk{p^{\flat-}_{ij}}{n}{p^{\flat+}_{ij}}}}\bmk{p^{\flat-}_{ij}}{\sigma^\mu}{p^{\flat+}_{ij}}\nonumber\\
a^\mu_-~&=&~a^{\mu*}_+\, .
 \end{eqnarray}
We've constructed these vectors so that $(a_++a_-)\cdot n=0$. Additionally, we have $a_\pm\cdot p_i=a_\pm\cdot p_j=0$. To see this, note that
  \begin{eqnarray}
a_+\cdot p_i~\sim~\bmk{p^{\flat-}_{ij}}{\,i\,}{p^{\flat+}_{ij}}\, ,
 \end{eqnarray}
 and since $p_i$ is a linear combination of $p^{\flat+}_{ij}$ and $p^{\flat-}_{ij}$ the whole expression is zero by the Dirac equation. By similar arguments $a_\pm\cdot p_i=a_\pm\cdot p_j=0$.
 \\\quad\\
Finally, we reproduce Zwanziger's Eq. (5.9):
 \begin{eqnarray}
a^\mu_+a^{\nu*}_+~&=&~\bmk{p^{\flat-}_{ij}}{\sigma^\mu}{p^{\flat+}_{ij}}\bmk{p^{\flat+}_{ij}}{\sigma^\nu}{p^{\flat-}_{ij}}\, \nonumber\\
a^\mu_-a^{\nu*}_-~&=&~\bmk{p^{\flat+}_{ij}}{\sigma^\mu}{p^{\flat-}_{ij}}\bmk{p^{\flat-}_{ij}}{\sigma^\nu}{p^{\flat+}_{ij}}\,.
 \end{eqnarray}
Using the identity
  \begin{eqnarray}\label{eq:3s}
\bmk{v}{\sigma^\mu}{u}\bmk{u}{\sigma^\nu}{v}~&=&~\frac{1}{v\cdot u}\left[
v^{\mu} u^{\nu}\,+\,u^{\mu} v^{\nu}\,-\,\left(v\cdot u\right)\,g^{\mu\nu}+i\,\epsilon^{\mu\nu\rho\sigma}\,v_\mu u_\rho\right]\, ,\nonumber\\
 \end{eqnarray}
 valid for any null vector $u,\,v$, we have
 \begin{eqnarray}
a^\mu_\pm \,a^{\nu*}_\pm~&=&~\frac{1}{(p^{\flat+}_{ij}\cdot p^{\flat-}_{ij})}\,\left[p^{\mu;\flat+}_{ij} p^{\nu;\flat-}_{ij}\,+\,p^{\mu;\flat-}_{ij} p^{\nu;\flat+}_{ij}\,-\,\left(p^{\flat+}_{ij}\cdot p^{\flat-}_{ij}\right)\,g^{\mu\nu}\,\mp\,i\,\epsilon^{\mu\nu\rho\sigma}\,p^{\flat+}_{\mu;ij} p^{\flat-}_{\rho;ij}\right]\, ,\nonumber\\
 \end{eqnarray}
 or explicitly
  \begin{eqnarray}
a^\mu_\pm\,a^{\nu*}_\pm~&=&~-g^{\mu\nu}+\,\frac{(p_i\cdot p_j)\,\left(p^{\mu}_i p^{\nu}_j\,+\,p^{\mu}_j p^{\nu}_i\right)-m^2_j\,p^\mu_i p^\nu_i\,-\,m^2_i\,p^\mu_j p^\nu_j}{(p_i\cdot p_j)^2-m^2_i\,m^2_j}\,-\nonumber\\
&&\mp\frac{i\,\epsilon^{\mu\nu\rho\sigma}\,p_{\mu;i} p_{\nu;j}}{\sqrt{(p_i\cdot p_j)^2-m^2_i\,m^2_j}}\, .
 \end{eqnarray}
 This is exactly Zwanziger's Eq. (5.9). Contracting this with $g^{\mu\nu}$, we see that
   \begin{eqnarray}
-\frac{1}{2}(a_\pm \cdot a^*_\pm)~=~1\, ,
 \end{eqnarray}
 and so $\hat{\epsilon}^\mu\equiv\frac{i}{2}(a^\mu_++a^{\mu*}_+)$ and $\hat{\zeta}^\mu\equiv\frac{1}{2}(a^\mu_+-a^{\mu*}_+)$ are two orthonormal vectors, orthogonal to $p^\mu_{i,j}$. By definition $\hat{\epsilon}\cdot n=0$.
 
 To show the LG covariance of $a^\mu_\pm$, we follow Zwanziger's argument. We note that 
\begin{eqnarray}
a^\mu_\pm(\Lambda p_i,\Lambda p_j,n)~=~\Lambda^\mu_{~\nu}\,a^{\nu}_\pm(p_i,p_j,\Lambda^{-1}n)\, .
 \end{eqnarray}
As we Lorentz transform, $n$, $\hat{\epsilon}$ remains in the plane orthogonal to $p_{i,j}$ and so is rotated by the angle $\phi_{ij}$ such that
\begin{eqnarray}
\cos \phi_{ij}~=~\hat{\epsilon}(\Lambda^{-1}n)\cdot \hat{\epsilon}(n)\, .
 \end{eqnarray}
Since $\hat{\zeta}\cdot \hat{\epsilon}=0$ and is also in the $\hat{\zeta}$ same plane orthogonal to $p_{i,j}$, it is rotated by the same angle. But since $a^\mu_\pm=i\hat{\epsilon}^\mu\pm \hat{\zeta}^\mu$, this rotation amounts to a phase factor $\exp(\pm i\phi_{ij})$ for $a^\mu_\pm$.
Summing up, we have
\begin{eqnarray}
a^\mu_\pm(\Lambda p_i,\Lambda p_j,n)~=~\Lambda^\mu_{~\nu}\,a^{\nu}_\pm(p_i,p_j,n)\exp(\pm i\phi_{ij})\, .
 \end{eqnarray}
The last thing to show is that the angle $\phi_{ij}$ is the same LG angle as in Eq.~(\ref{eq:pole-charge_Lorentz_transf_spinless}). But Zwanziger shows that we can always fix the $U(1)$ ambiguity in the definition of $L(p_i,p_j)$ such that:
\begin{eqnarray}
L(p_i,p_j)^\mu_2~=~\hat{\epsilon}^\mu\, ,
 \end{eqnarray}
and consequently the LG rotation angle is exactly the rotation angle of $\hat{\epsilon}$.

\section{Comparison of amplitude formalism to QM calculations}\label{sec:kazama-compare}
Here we show that Eq.~(\ref{eq:kazj}) exactly reproduces the angular dependence of the higher partial amplitudes in \cite{Kazama:1976fm}. Starting from their partial amplitude
\begin{eqnarray}\label{eq:kazjtheirs}
 &&S^{J}_{f\rightarrow f}~=~S^{J}_{\bar{f}\rightarrow \bar{f}}~=~\nonumber\\[10pt]
&&\mathcal{N}~e^{-i\pi\mu}\,\frac{\mu}{\cos(\theta_c/2)}\,\left[\sqrt{\frac{4\pi}{2j}}\,\,{}_qY_{j-\frac{1}{2},-q}(-\Omega_c)\,-\,\sqrt{\frac{4\pi}{2j+2}}\,\,{}_qY_{j+\frac{1}{2},-q}(-\Omega_c)\,\right]\, ,
\end{eqnarray}
where $-\Omega_c=\left(\pi-\theta_c,-\phi_c\right)$ and $\mu\equiv\sqrt{(J+\frac{1}{2})^2-q^2}$. 
and using the relation Eq.~(\ref{eq:DYqlmb}) between the ${}_qY_{lm}$ and Wigner $\mathcal{D}$-matrices, we can cast it in the form 
\begin{eqnarray}\label{eq:kazjtheirs}
 &&S^{J}_{f\rightarrow f}~=~S^{J}_{\bar{f}\rightarrow \bar{f}}~=~\mathcal{N}\,e^{-i\pi\mu}~\frac{\mu}{\cos(\theta_c/2)}\,\left[\mathcal{D}^{J-\frac{1}{2}\,*}_{q,-q}\left(\Omega_c\right)+\mathcal{D}^{J+\frac{1}{2}\,*}_{q,-q}\left(\Omega_c\right)\,\right]\, .~~~~~~~~~~~
\end{eqnarray}
Finally we can use $\mathcal{D}$-matrix identities in sec 4.8.2 of \cite{Varshalovich:1988ye} to transform this expression to
\begin{eqnarray}\label{eq:kazjtheirs2}
 &&S^{J}_{f\rightarrow f}~=~S^{J}_{\bar{f}\rightarrow \bar{f}}~=\nonumber\\[10pt]
&&\mathcal{N}~(2J+1)\,e^{-i\pi\mu}~\mathcal{D}^{J*}_{q-\frac{1}{2},-q-\frac{1}{2}}(\Omega_c)~=~\mathcal{N}~(2J+1)\,e^{-i\pi\mu}~\mathcal{D}^{J*}_{q+\frac{1}{2},-q+\frac{1}{2}}(\Omega_c)\, .
\end{eqnarray}
Comparing this to the result obtained in our amplitude formalism,Eq.~(\ref{eq:kazj}), implies that
\begin{eqnarray}\label{eq:mj}
\mathcal{M}^J_{\pm\frac{1}{2},\pm\frac{1}{2}}~=~e^{-i\pi\mu}\,.
\end{eqnarray}
where $\mu=\sqrt{\left(J+\frac{1}{2}\right)^2-q^2}$. Combining this expression with the unitarity condition leads to
\begin{eqnarray}\label{eq:kazjtheirs4}
\left\vert \mathcal{M}^J_{\pm\frac{1}{2},\mp\frac{1}{2}}\right\vert^2~=~1-\left\vert \mathcal{M}^J_{\pm\frac{1}{2},\pm\frac{1}{2}} \right\vert^2~=~0\, ,
\end{eqnarray}
for helicity-flip $J>|g|-\frac{1}{2}$ processes in an agreement with the explicit calculation in \cite{Kazama:1976fm}.

\bibliographystyle{JHEP}
\bibliography{AMP}{}

\providecommand{\href}[2]{#2}\begingroup\raggedright\begin{thebibliography}{10}

\bibitem{Wigner:1939cj}
E.~P. Wigner, \emph{{On Unitary Representations of the Inhomogeneous Lorentz
  Group}}, \href{http://dx.doi.org/10.2307/1968551}{\emph{Annals Math.}
  {\bfseries 40} (1939) 149--204}.

\bibitem{Weinberg:1995mt}
S.~Weinberg, \emph{{The Quantum theory of fields. Vol. 1: Foundations}}.
\newblock Cambridge University Press, 2005.

\bibitem{Zwanziger:1972sx}
D.~Zwanziger, \emph{Angular distributions and a selection rule in charge-pole
  reactions}, \href{http://dx.doi.org/10.1103/PhysRevD.6.458}{\emph{Phys. Rev.
  D} {\bfseries 6} (Jul, 1972) 458--470}.

\bibitem{Dirac:1948um}
P.~A. Dirac, \emph{{The Theory of magnetic poles}},
  \href{http://dx.doi.org/10.1103/PhysRev.74.817}{\emph{Phys. Rev.} {\bfseries
  74} (1948) 817--830}.

\bibitem{Zwanziger:1970hk}
D.~Zwanziger, \emph{{Local Lagrangian quantum field theory of electric and
  magnetic charges}},
  \href{http://dx.doi.org/10.1103/PhysRevD.3.880}{\emph{Phys. Rev. D}
  {\bfseries 3} (1971) 880}.

\bibitem{Weinberg:1965rz}
S.~Weinberg, \emph{{Photons and gravitons in perturbation theory: Derivation of
  Maxwell's and Einstein's equations}},
  \href{http://dx.doi.org/10.1103/PhysRev.138.B988}{\emph{Phys. Rev.}
  {\bfseries 138} (1965) B988--B1002}.

\bibitem{Terning:2018udc}
J.~Terning and C.~B. Verhaaren, \emph{{Resolving the Weinberg Paradox with
  Topology}}, \href{http://dx.doi.org/10.1007/JHEP03(2019)177}{\emph{JHEP}
  {\bfseries 03} (2019) 177},
  [\href{https://arxiv.org/abs/1809.05102}{{\ttfamily 1809.05102}}].

\bibitem{Dirac:1931kp}
P.~A.~M. Dirac, \emph{{Quantised singularities in the electromagnetic field,}},
  \href{http://dx.doi.org/10.1098/rspa.1931.0130}{\emph{Proc. Roy. Soc. Lond.}
  {\bfseries A133} (1931) 60--72}.

\bibitem{Laperashvili:1999pu}
L.~Laperashvili and H.~B. Nielsen, \emph{{Dirac relation and renormalization
  group equations for electric and magnetic fine structure constants}},
  \href{http://dx.doi.org/10.1142/S0217732399002935}{\emph{Mod. Phys. Lett. A}
  {\bfseries 14} (1999) 2797},
  [\href{https://arxiv.org/abs/hep-th/9910101}{{\ttfamily hep-th/9910101}}].

\bibitem{Gamberg:1999hq}
L.~P. Gamberg and K.~A. Milton, \emph{{Dual quantum electrodynamics: Dyon-dyon
  and charge monopole scattering in a high-energy approximation}},
  \href{http://dx.doi.org/10.1103/PhysRevD.61.075013}{\emph{Phys. Rev. D}
  {\bfseries 61} (2000) 075013},
  [\href{https://arxiv.org/abs/hep-ph/9910526}{{\ttfamily hep-ph/9910526}}].

\bibitem{Bruemmer:2009ky}
F.~Brummer, J.~Jaeckel and V.~V. Khoze, \emph{{Magnetic Mixing: Electric
  Minicharges from Magnetic Monopoles}},
  \href{http://dx.doi.org/10.1088/1126-6708/2009/06/037}{\emph{JHEP} {\bfseries
  06} (2009) 037}, [\href{https://arxiv.org/abs/0905.0633}{{\ttfamily
  0905.0633}}].

\bibitem{Csaki:2010rv}
C.~Csaki, Y.~Shirman and J.~Terning, \emph{{Anomaly Constraints on Monopoles
  and Dyons}}, \href{http://dx.doi.org/10.1103/PhysRevD.81.125028}{\emph{Phys.
  Rev. D} {\bfseries 81} (2010) 125028},
  [\href{https://arxiv.org/abs/1003.0448}{{\ttfamily 1003.0448}}].

\bibitem{Sanchez:2011mf}
C.~Gomez~Sanchez and B.~Holdom, \emph{{Monopoles, strings and dark matter}},
  \href{http://dx.doi.org/10.1103/PhysRevD.83.123524}{\emph{Phys. Rev. D}
  {\bfseries 83} (2011) 123524},
  [\href{https://arxiv.org/abs/1103.1632}{{\ttfamily 1103.1632}}].

\bibitem{Colwell:2015wna}
K.~Colwell and J.~Terning, \emph{{S-Duality and Helicity Amplitudes}},
  \href{http://dx.doi.org/10.1007/JHEP03(2016)068}{\emph{JHEP} {\bfseries 03}
  (2016) 068}, [\href{https://arxiv.org/abs/1510.07627}{{\ttfamily
  1510.07627}}].

\bibitem{Hook:2017vyc}
A.~Hook and J.~Huang, \emph{{Bounding millimagnetically charged particles with
  magnetars}}, \href{http://dx.doi.org/10.1103/PhysRevD.96.055010}{\emph{Phys.
  Rev. D} {\bfseries 96} (2017) 055010},
  [\href{https://arxiv.org/abs/1705.01107}{{\ttfamily 1705.01107}}].

\bibitem{Terning:2018lsv}
J.~Terning and C.~B. Verhaaren, \emph{{Dark Monopoles and $SL(2,\mathbb Z)$
  Duality}}, \href{http://dx.doi.org/10.1007/JHEP12(2018)123}{\emph{JHEP}
  {\bfseries 12} (2018) 123},
  [\href{https://arxiv.org/abs/1808.09459}{{\ttfamily 1808.09459}}].

\bibitem{Caron-Huot:2018ape}
S.~Caron-Huot and Z.~Zahraee, \emph{{Integrability of Black Hole Orbits in
  Maximal Supergravity}},
  \href{http://dx.doi.org/10.1007/JHEP07(2019)179}{\emph{JHEP} {\bfseries 07}
  (2019) 179}, [\href{https://arxiv.org/abs/1810.04694}{{\ttfamily
  1810.04694}}].

\bibitem{Huang:2019cja}
Y.-T. Huang, U.~Kol and D.~O'Connell, \emph{{Double copy of electric-magnetic
  duality}}, \href{http://dx.doi.org/10.1103/PhysRevD.102.046005}{\emph{Phys.
  Rev. D} {\bfseries 102} (2020) 046005},
  [\href{https://arxiv.org/abs/1911.06318}{{\ttfamily 1911.06318}}].

\bibitem{Moynihan:2020gxj}
N.~Moynihan and J.~Murugan, \emph{{On-Shell Electric-Magnetic Duality and the
  Dual Graviton}},  \href{https://arxiv.org/abs/2002.11085}{{\ttfamily
  2002.11085}}.

\bibitem{Arkani-Hamed:2017jhn}
N.~Arkani-Hamed, T.-C. Huang and Y.-t. Huang, \emph{{Scattering Amplitudes For
  All Masses and Spins}},  \href{https://arxiv.org/abs/1709.04891}{{\ttfamily
  1709.04891}}.

\bibitem{Kazama:1976fm}
Y.~Kazama, C.~N. Yang and A.~S. Goldhaber, \emph{{Scattering of a Dirac
  Particle with Charge Ze by a Fixed Magnetic Monopole}},
  \href{http://dx.doi.org/10.1103/PhysRevD.15.2287}{\emph{Phys. Rev.}
  {\bfseries D15} (1977) 2287--2299}.

\bibitem{Thomson}
J.~J. Thomson, \emph{On momentum in the electric field}, {\emph{Phil. Mag.}
  {\bfseries 8} (1904) 331}.

\bibitem{Schwinger:1969ib}
J.~S. Schwinger, \emph{{A Magnetic model of matter}},
  \href{http://dx.doi.org/10.1126/science.165.3895.757}{\emph{Science}
  {\bfseries 165} (1969) 757--761}.

\bibitem{Zwanziger:1969by}
D.~Zwanziger, \emph{{Exactly soluble nonrelativistic model of particles with
  both electric and magnetic charges}},
  \href{http://dx.doi.org/10.1103/PhysRev.176.1480}{\emph{Phys. Rev.}
  {\bfseries 176} (1968) 1480--1488}.

\bibitem{Lipkin:1969ck}
H.~Lipkin, W.~Weisberger and M.~Peshkin, \emph{{Magnetic charge quantization
  and angular momentum}},
  \href{http://dx.doi.org/10.1016/0003-4916(69)90279-6}{\emph{Annals Phys.}
  {\bfseries 53} (1969) 203--214}.

\bibitem{Wu:1976ge}
T.~T. Wu and C.~N. Yang, \emph{{Dirac Monopole Without Strings: Monopole
  Harmonics}},
  \href{http://dx.doi.org/10.1016/0550-3213(76)90143-7}{\emph{Nucl. Phys. B}
  {\bfseries 107} (1976) 365}.

\bibitem{Schuster:2014hca}
P.~Schuster and N.~Toro, \emph{{Continuous-spin particle field theory with
  helicity correspondence}},
  \href{http://dx.doi.org/10.1103/PhysRevD.91.025023}{\emph{Phys. Rev. D}
  {\bfseries 91} (2015) 025023},
  [\href{https://arxiv.org/abs/1404.0675}{{\ttfamily 1404.0675}}].

\bibitem{Elvang:2013cua}
H.~Elvang and Y.-t. Huang, \emph{{Scattering Amplitudes}},
  \href{https://arxiv.org/abs/1308.1697}{{\ttfamily 1308.1697}}.

\bibitem{Henn:2014yza}
J.~M. Henn and J.~C. Plefka, \emph{{Scattering Amplitudes in Gauge Theories}},
  vol.~883.
\newblock Springer, Berlin, 2014,
  \href{http://dx.doi.org/10.1007/978-3-642-54022-6}{10.1007/978-3-642-54022-6}.

\bibitem{Cheung:2017pzi}
C.~Cheung, \emph{{TASI Lectures on Scattering Amplitudes}}.
\newblock \href{https://arxiv.org/abs/1708.03872}{{\ttfamily 1708.03872}}.

\bibitem{Kosower:2004yz}
D.~A. Kosower, \emph{{Next-to-maximal helicity violating amplitudes in gauge
  theory}}, \href{http://dx.doi.org/10.1103/PhysRevD.71.045007}{\emph{Phys.
  Rev. D} {\bfseries 71} (2005) 045007},
  [\href{https://arxiv.org/abs/hep-th/0406175}{{\ttfamily hep-th/0406175}}].

\bibitem{Ossola:2006us}
G.~Ossola, C.~G. Papadopoulos and R.~Pittau, \emph{{Reducing full one-loop
  amplitudes to scalar integrals at the integrand level}},
  \href{http://dx.doi.org/10.1016/j.nuclphysb.2006.11.012}{\emph{Nucl. Phys. B}
  {\bfseries 763} (2007) 147--169},
  [\href{https://arxiv.org/abs/hep-ph/0609007}{{\ttfamily hep-ph/0609007}}].

\bibitem{Bern:1994cg}
Z.~Bern, L.~J. Dixon, D.~C. Dunbar and D.~A. Kosower, \emph{{Fusing gauge
  theory tree amplitudes into loop amplitudes}},
  \href{http://dx.doi.org/10.1016/0550-3213(94)00488-Z}{\emph{Nucl. Phys. B}
  {\bfseries 435} (1995) 59--101},
  [\href{https://arxiv.org/abs/hep-ph/9409265}{{\ttfamily hep-ph/9409265}}].

\bibitem{Forde:2007mi}
D.~Forde, \emph{{Direct extraction of one-loop integral coefficients}},
  \href{http://dx.doi.org/10.1103/PhysRevD.75.125019}{\emph{Phys. Rev. D}
  {\bfseries 75} (2007) 125019},
  [\href{https://arxiv.org/abs/0704.1835}{{\ttfamily 0704.1835}}].

\bibitem{Jiang:2020sdh}
M.~Jiang, J.~Shu, M.-L. Xiao and Y.-H. Zheng, \emph{{New Selection Rules from
  Angular Momentum Conservation}},
  \href{https://arxiv.org/abs/2001.04481}{{\ttfamily 2001.04481}}.

\bibitem{Durieux:2020gip}
G.~Durieux, T.~Kitahara, C.~S. Machado, Y.~Shadmi and Y.~Weiss,
  \emph{{Constructing massive on-shell contact terms}},
  \href{https://arxiv.org/abs/2008.09652}{{\ttfamily 2008.09652}}.

\bibitem{Witten:2003nn}
E.~Witten, \emph{{Perturbative gauge theory as a string theory in twistor
  space}}, \href{http://dx.doi.org/10.1007/s00220-004-1187-3}{\emph{Commun.
  Math. Phys.} {\bfseries 252} (2004) 189--258},
  [\href{https://arxiv.org/abs/hep-th/0312171}{{\ttfamily hep-th/0312171}}].

\bibitem{Conde:2016izb}
E.~Conde, E.~Joung and K.~Mkrtchyan, \emph{{Spinor-Helicity Three-Point
  Amplitudes from Local Cubic Interactions}},
  \href{http://dx.doi.org/10.1007/JHEP08(2016)040}{\emph{JHEP} {\bfseries 08}
  (2016) 040}, [\href{https://arxiv.org/abs/1605.07402}{{\ttfamily
  1605.07402}}].

\bibitem{Guevara:2018wpp}
A.~Guevara, A.~Ochirov and J.~Vines, \emph{{Scattering of Spinning Black Holes
  from Exponentiated Soft Factors}},
  \href{http://dx.doi.org/10.1007/JHEP09(2019)056}{\emph{JHEP} {\bfseries 09}
  (2019) 056}, [\href{https://arxiv.org/abs/1812.06895}{{\ttfamily
  1812.06895}}].

\bibitem{Rubakov:1981rg}
V.~A. Rubakov, \emph{{Superheavy Magnetic Monopoles and Proton Decay}},
  {\emph{JETP Lett.} {\bfseries 33} (1981) 644--646}.

\bibitem{Callan:1982ac}
C.~G. Callan, Jr., \emph{{Monopole Catalysis of Baryon Decay}},
  \href{http://dx.doi.org/10.1016/0550-3213(83)90677-6}{\emph{Nucl. Phys.}
  {\bfseries B212} (1983) 391--400}.

\bibitem{Varshalovich:1988ye}
D.~Varshalovich, A.~Moskalev and V.~Khersonsky, \emph{{Quantum Theory of
  Angular Momentum: Irreducible Tensors, Spherical Harmonics, Vector Coupling
  Coefficients, 3nj Symbols}}.
\newblock World Scientific, Singapore, 1988.

\bibitem{Schwinger:1976fr}
J.~S. Schwinger, K.~A. Milton, W.-y. Tsai, J.~DeRaad, Lester~L. and D.~C.
  Clark, \emph{{Nonrelativistic Dyon-Dyon Scattering}},
  \href{http://dx.doi.org/10.1016/0003-4916(76)90020-8}{\emph{Annals Phys.}
  {\bfseries 101} (1976) 451}.

\bibitem{Boulware:1976tv}
D.~G. Boulware, L.~S. Brown, R.~N. Cahn, S.~Ellis and C.-k. Lee,
  \emph{{Scattering on Magnetic Charge}},
  \href{http://dx.doi.org/10.1103/PhysRevD.14.2708}{\emph{Phys. Rev. D}
  {\bfseries 14} (1976) 2708}.

\bibitem{Pilkuhn:1979ps}
H.~Pilkuhn, \emph{{Relativistic Particle Physics}}.
\newblock 1, 1979.

\bibitem{Griest:1989wd}
K.~Griest and M.~Kamionkowski, \emph{{Unitarity Limits on the Mass and Radius
  of Dark Matter Particles}},
  \href{http://dx.doi.org/10.1103/PhysRevLett.64.615}{\emph{Phys. Rev. Lett.}
  {\bfseries 64} (1990) 615}.

\bibitem{Terningprogress}
J.~Terning and C.~B. Verhaaren, \emph{{work in progress}}, .

\bibitem{Hannesdottir:2019opa}
H.~Hannesdottir and M.~D. Schwartz, \emph{{$S$ -Matrix for massless
  particles}}, \href{http://dx.doi.org/10.1103/PhysRevD.101.105001}{\emph{Phys.
  Rev. D} {\bfseries 101} (2020) 105001},
  [\href{https://arxiv.org/abs/1911.06821}{{\ttfamily 1911.06821}}].

\bibitem{Kol:2020ucd}
U.~Kol and M.~Porrati, \emph{{Gravitational Wu-Yang Monopoles}},
  \href{http://dx.doi.org/10.1103/PhysRevD.101.126009}{\emph{Phys. Rev. D}
  {\bfseries 101} (2020) 126009},
  [\href{https://arxiv.org/abs/2003.09054}{{\ttfamily 2003.09054}}].

\end{thebibliography}\endgroup

\end{document}